\documentclass[11pt]{article}
\usepackage{amssymb}          
\usepackage{graphicx}

\makeatletter                      
 
 \@addtoreset{equation}{section}
\makeatother

\newcommand{\non}{\nonumber}
\newcommand{\noi}{\noindent}

\newcommand{\eqref}[1]{(\ref{#1})}
\newcommand{\bea}{\begin{eqnarray}}
\newcommand{\ena}{\end{eqnarray}}
\newcommand{\be}{\begin{eqnarray*}}
\newcommand{\en}{\end{eqnarray*}}
\newcommand{\lb}[1]{\label{#1}}

\newcommand{\dstyle}{\displaystyle}
\newcommand{\sstyle}{\scriptstyle}

\newcommand{\C}{{\cal C}}
\newcommand{\F}{{\cal F}}
\newcommand{\M}{{\cal M}}
\renewcommand{\H}{{\cal H}}
\newcommand{\id}{{\rm id}}
\newcommand{\tr}{{\rm tr}}

\newcommand{\rs}{{r^*}}
\newcommand{\ve}{{\varepsilon}}
\newcommand{\eps}{{\epsilon}}
\newcommand{\bPhi}{{{\bar\Phi}}}
\newcommand{\ka}{{\kappa}}
\newcommand{\z}{{\zeta}}

\newcommand{\sn}{\mbox{sn}}

\newcommand{\br}[1]{{\langle #1 \rangle}} 
\newcommand{\bra}[1]{\langle #1 |}        
\newcommand{\ket}[1]{{| #1 \rangle}}      
\newcommand{\dbr}[1]{{\langle\!\langle #1 \rangle\!\rangle}} 
\newcommand{\lbs}[2]{ {_{B}^{\ c}}\langle {#1},{#2}|}
\newcommand{\rbs}[2]{|{#1},{#2}\rangle_B^c}

\newcommand{\BW}[5]
{\!\left(\begin{array}{cc}#1 & #2 \cr #3 & #4 \cr\end{array}
\Biggl| #5\right)}
\newcommand{\Kmatrix}[4]
{\!\left(#1\begin{array}{c}#2\\#3\end{array}\Bigl|#4\right)}
\newcommand{\LL}[5]
{\!\left[\begin{array}{cc}#1 & #2 \cr #3 & #4 \cr\end{array}
\Biggl| #5\right]}

\begin{document}

\font\csc=cmcsc10 scaled\magstep1

\vspace*{21mm}
\begin{center}
{\Large\bf Correlation functions of the XYZ model\\ with a boundary} 
\end{center}
\vspace{11mm}

\begin{center}
{\csc
Yuji HARA\\Institute of Physics,\\ Graduate School of Arts and Sciences, \\
University of Tokyo, Tokyo 153-8902, Japan
}
\end{center}

\begin{abstract}
Integral formulae for the correlation functions of the XYZ model with a boundary are calculated by mapping the model to the bosonized boundary SOS model. The boundary $K$-matrix considered here coincides with the known general solution of the boundary Yang-Baxter equation. For the case of diagonal $K$-matrix, our formulae reproduce the one-point function previously obtained by solving boundary version of quantum Knizhnik-Zamolodchikov equation.
\end{abstract}

\renewcommand{\thefootnote}{\fnsymbol{footnote}}
\footnote[0]{e-mail:ss77070@komaba.ecc.u-tokyo.ac.jp}
%
%



\setcounter{section}{0}
\setcounter{equation}{0}
\section{Introduction}
\lb{sec:Intro}

The XYZ model, the eight-vertex model and the ABF model are the representative examples of solvable lattice models in statistical mechanics \cite{Bax}.  In addition to the exact results e.g. free energy, one-point function, relations with many areas of mathematics and physics have been studied: conformal field theory, integrable quantum field theory, theory of Lie algebras and its representation theory. In the last years, the boundary problems of the solvable lattice models have been investigated intensively. Under a special boundary condition the models with boundaries are also solvable \cite{Skl}. In this case the boundary condition is described by the so called $K$-matrix and the integrability of the model is ensured by the boundary version of the Yang-Baxter equation which includes the $K$-matrix in addition to the $R$-matrix.

There are several ways to solve the models. Among them the vertex operator approach is the one which allows us to get deep insights on the symmetry of the model and to obtain integral formulae for the correlation functions. This approach is first launched on the XXZ model \cite{JM}. In this case the symmetry of the model is described with the quantum affine algebra $U_q\bigl(\widehat{\mathfrak{sl}}_2\bigr)$. Representation theoretical counter parts of the space of the states, the transfer matrix and the creation operator of the elementary excitation are the level one higest modules, type I vertex operators and type II vertex operators of $U_q\bigl(\widehat{\mathfrak{sl}}_2\bigr)$ respectively. By bosonizing the algebra, the modules and the vertex operators, the hamiltonian is diagonalized and the formulae of the correlation functions are obtained. The same recipe is applied to the ABF model and its symmetry is described by the deformed Virasoro algebra \cite{LuP}. As for the boundary problems, the half-infinte XXZ model with a diagonal $K$-matrix is studied by this approach in \cite{bdry XXZ}. The vacuum states are constructed in Fock spaces and the correlation functions are obtained. For the boundary ABF model, daigonalization of the transfer matrix is carried through in \cite{MW}.

On the other hand, for the XYZ model, until recently the bosonization was not known. Diffence equations of quantum Knizhnik-Zamolodchikov types are used to get the correlation functions. The bulk case is discussed in \cite{bulk qKZ} and the boundary case is in \cite{bdry qKZ} for a diagonal $K$-matrix. The bosonization is achieved in \cite{LaP}, \cite{LaP1} by mapping the model to the ABF model without restriction (the SOS model) through the face-vertex correspondence. The correlation functions of the XYZ model are expressed with that of the SOS model with non local insertion called the tail operator.  The Baxter-Kelland formula for the one-point function is reproduced as a special case.

In this paper, we apply the bosonization scheme of \cite{LaP} to the boundary problem of the XYZ model. The $K$-matrix discussed here is obtained from the diagonal $K$-matrix of the ABF model through face-vertex correspondence. It has off-diagonal elements and coincides with that of \cite{IK} which is a general solution of the boundary Yang-Baxter equation. We obtain the integral formulae for the correlation functions for this general boundary condition. And as a special case of the formulae the results of \cite{bdry qKZ} and \cite{bdry XXZ} are reproduced .

In Sec.\ref{sec:XYZ} we formulate the boundary XYZ model. In Sec.\ref{sec:SOS}, we recall the results of \cite{MW}. Modification to the unrestricted case is almost trivial. Sec.\ref{sec:corr-fn} is the main part. After discussing the face-vertex correspondence, we give the formulae for the correlation functions. In Appendix \ref{appdx:IK}, we detail the correspondence of our $K$-matrix and the one of \cite{IK}.


\setcounter{section}{1}
\setcounter{equation}{0}
\section{The boundary XYZ model}
\lb{sec:XYZ}


\subsection{The bulk weights and lattice vertex operators}
First we fix the convention of the $R$-matirx (Fig.\ref{fig:R-K}). Originally the $R$-matrix is found by Baxter as the local Boltzman weight of the eight-vertex model \cite{Bax}. Here we follow \cite{LaP} in which elements $R(u)_{++}^{--}=R(u)_{--}^{++}$ are negated as compared to the one in \cite{Bax}. The elements are defined as
\bea
&&
a(u)=R(u)_{++}^{++}=R(u)_{--}^{--}
=-i\ka (u) R_0(u)\,
\textstyle{
\vartheta_4\!\left(i{\epsilon\over\pi};i{2\epsilon r\over\pi}\right)
\vartheta_4\!\left(i{\epsilon\over\pi}u;i{2\epsilon r\over\pi}\right)
\vartheta_1\!\left(i{\epsilon\over\pi}(1-u);i{2\epsilon r\over\pi}\right),
}
\non\\
&&
b(u)=R(u)_{+-}^{+-}=R(u)_{-+}^{-+}
=-i\ka (u) R_0(u)\,
\textstyle{
\vartheta_4\!\left(i{\epsilon\over\pi};i{2\epsilon r\over\pi}\right)
\vartheta_1\!\left(i{\epsilon\over\pi}u;i{2\epsilon r\over\pi}\right)
\vartheta_4\!\left(i{\epsilon\over\pi}(1-u);i{2\epsilon r\over\pi}\right),
}
\non\\
&&
c(u)=R(u)_{+-}^{-+}=R(u)_{-+}^{+-}
=-i\ka (u) R_0(u)\,
\textstyle{
\vartheta_1\!\left(i{\epsilon\over\pi};i{2\epsilon r\over\pi}\right)
\vartheta_4\!\left(i{\epsilon\over\pi}u;i{2\epsilon r\over\pi}\right)
\vartheta_4\!\left(i{\epsilon\over\pi}(1-u);i{2\epsilon r\over\pi}\right),
}
\non\\
&&
d(u)=R(u)_{++}^{--}=R(u)_{--}^{++}
=-i\ka (u) R_0(u)\,
\textstyle{
\vartheta_1\!\left(i{\epsilon\over\pi};i{2\epsilon r\over\pi}\right)
\vartheta_1\!\left(i{\epsilon\over\pi}u;i{2\epsilon r\over\pi}\right)
\vartheta_1\!\left(i{\epsilon\over\pi}(1-u);i{2\epsilon r\over\pi}\right)
.}\non\\
&&
{}
\ena
Notation of theta functions are given in Appendix \ref{appdx:fn}.
We use the parameters $\eps$ and $r$ and consider the so called principal regime
\bea
\eps>0,\quad r>1,\quad -1<u<1.
\lb{eqn:pr-reg}
\ena
We also use the follwing parameters
\be
x=e^{-\epsilon},\qquad p=x^{2r},\qquad \z=x^{2u}.
\en
The common factor $\ka (u)R_0(u)$ is so chosen that
the partition function per site is equal to one
\bea
&&
\ka (u)
=\z^{-{r-1\over 2r}}x^{1-{r\over2}}
(x^{2r};x^{2r})_\infty^{-2}(x^{4r};x^{4r})_\infty^{-1}
(x^2\z^{-1};x^{2r})_\infty^{-1}(x^{2r-2}\z;x^{2r})_\infty^{-1},
\\[2mm]
&&
R_0(u)=\z^{\frac{r-1}{2r}}\frac{\rho(\z)}{\rho(\z^{-1})},
\lb{eqn:norm-R}\\
&&
\rho(z)={ \{x^4 z\}  \{x^{2r} z\} \over 
 \{x^2 z\}  \{x^{2r+2} z\} },\\
&&
\{z\}=(z;x^4,x^{2r})_\infty,\\
&&
(z;p_1,\ldots,p_N)_\infty
=\prod_{n_1,\ldots,n_N=0}^\infty(1-zp_1^{n_1}\ldots p_N^{n_N}).
\ena
The $R$-matrix satisfies the relations\\
\noi
(i) Yang-Baxter equation
\bea
R_{12}(u_1-u_2)R_{13}(u_1-u_3)R_{23}(u_2-u_3)=R_{23}(u_2-u_3)R_{13}(u_1-u_3)R_{12}(u_1-u_2),
\ena
(ii) Unitarity relation
\bea
R_{12}(u_1-u_2)R_{21}(u_2-u_1)=\id,
\ena
(iii) Crossing relation
\bea
R^{\ve_1\ve_2}_{\ve_1'\ve_2'}(1-u)=R^{\ve_2\:-\ve_1'}_{\ve_2'\:-\ve_1}(u).
\ena

We will call the half infinite transfer matrix $\phi_\ve(u)$ \lq\lq lattice vertex operator of vertex type'' and $\phi^*_\ve(u)$ its \lq\lq dual'' (Fig.\ref{fig:VO-v}) \cite{bulk qKZ}. A vertex-path is a semi-infinite sequence $(\ldots,p(2),p(1)),\;v(n)\in\{+,-\}$. We denote by $\H^i,\;\;i=0\mbox{ or }1$ the eigenspace of the corner transfer matrix \cite{Bax} in the NE quadrant spanned by paths such that $p(n)=(-1)^{n-i}\mbox{ for }n\gg 1$. Then the lattice vertex operators act on these spaces as
\bea
\phi_\ve(u),\;\phi^*_\ve(u):\H^i\to \H^{1-i}.
\ena 
All these are unrigorous picture but for the XXZ model, the model has the symmetry described by the affine quantum group $U_q\bigl(\widehat{\mathfrak{sl}}_2\bigr)$ and $\H^i$ are identified with the level one higest weight modules of $U_q\bigl(\widehat{\mathfrak{sl}}_2\bigr)$ \cite{JM}. And lattice vertex operators are identified with the intertwiners for these modules.
For the eight-vertex model, the elliptic affine quantum group $\mathcal{A}_{q,p}\bigl(\widehat{\mathfrak{sl}}_2\bigr)$ is expected to play the same role \cite{Aqp}.


\subsection{The boundary XYZ model}

The boundary of the model is formulated with a $K$-matrix which satisfies the following equations \cite{Skl} (Fig.\ref{fig:R-K}).\\
\noi
(iv) Boundary Yang-Baxter equation
\bea
&&
K_2(u_2)R_{21}(u_1+u_2)K_1(u_1)R_{12}(u_1-u_2)
=R_{21}(u_1-u_2)K_1(u_1)R_{12}(u_1+u_2)K_2(u_2),\non\\
&&\lb{eqn:bdryYBE}
\ena
(v) Boundary unitarity relation
\bea
K(u)K(-u)=\hbox{\rm id},
\lb{bdryU}
\ena
(vi) Boundary crossing relation
\be
K^b_a(1-u)=\sum_{a',b'}R^{b,\,-a'}_{-b'\,a}(2-2u)K^{a'}_{b'}(u).
\lb{eqn:bdryX}
\en
Explicit form of $K$-matrix will be given in Sec.\ref{subsec:fv-K}. 


Then the transfer matrix for the boundary eight-vertex model $T_B(u)$ is defined as
\bea
T_B(u)=\sum_{\ve,\ve'}\phi^*_{\ve'}(-u)K^\ve_{\ve'}(u)\phi_\ve(u),
\ena
The boundary Yang-Baxter equation ensures the integrability of this model
\bea
[T_B(u),T_B(v)]=0.
\ena

The Hamiltonian of the XYZ model with a boundary is defined as
\bea 
H_B=-\frac{1}{2}\sum_{k=1}^{\infty}\Bigl((1-\Gamma)
\sigma^x_{k+1}\sigma^x_{k}+(1+\Gamma)
\sigma^y_{k+1}\sigma^y_{k}+\Delta\sigma^z_{k+1}\sigma^z_{k}\Bigr)
+h_x\sigma^x_1+h_y\sigma^y_1+h_z\sigma^z_1,
\lb{eqn:hamiltonian}
\ena
and it is connected to the boundary eight-vertex model as
\bea
H^{ren}_B=-\frac{\pi {\rm snh}(2\eps K/\pi,k)}{4K}
\z\frac{d}{d\z}T_B(u)\Bigl|_{\z=1},
\lb{eqn:Hamil-ren}
\ena
where
\bea
&&
\Gamma=k\,{\rm snh}^2(2\eps K/\pi,k),\quad 
\Delta=-{\rm cnh}(2\eps K/\pi,k){\rm dnh}(2\eps K/\pi,k),\\[2mm]
&&
h_x\sigma^x_1+h_y\sigma^y_1=h_+\sigma^+_1+h_-\sigma^-_1,\quad
h_+=\z\frac{d}{d\z}K^-_+(u)\Bigl|_{\z=1},\quad h_-=\z\frac{d}{d\z}K^+_-(u)\Bigl|_{\z=1},\\[2mm]
&&
h_z=\z\frac{d}{d\z}\left(K^+_+(u)-K^-_-(u)\right)\Bigl|_{\z=1}.
\ena
See Appendix \ref{appdx:fn} for elliptic functions snh, cnh, dnh and the elliptic modulus $k$, the half-period magnitude $K$.

We assume that the boundary transfer matrix $T_B(u)$ acts on the well defined infinite dimensional subspace of $\ldots\otimes\mathbb{C}^2\otimes\mathbb{C}^2\otimes\mathbb{C}^2$ and denote it by $\H^i_B\;\;i=0\mbox{ or }1$. The space $\H^i_B$ is the eigenspace of $T_B(u)$ spanned by paths such that $p(n)=(-1)^{n-i}\mbox{ for }n\gg 1$. Then
\bea
T_B(u):\H^i_B\to\H^i_B.
\ena
Contrary to the bulk problems mentioned in the previous subsection, affine quantum algebras are not the symmetry of the models anymore and the representation theoretical meaning of the space $\H^i_B$ is not clarified yet even for the XXZ model \cite{bdry XXZ}.

We assume existence of the lowest energy state in each sector $\ket{i}_B\in\H^i_B$ at least for the regions
\be
&&
\Gamma=0,\Delta\to-\infty,\quad h_z:\mbox{arbitrary},\; h_x,h_y:\mbox{small},\\
\mbox{and    }&&\\
&&
\Gamma,\Delta:\mbox{arbitrary},\quad h_z:\mbox{small},\quad h_x,h_y:\mbox{small}.
\en
\noi
When $\Gamma=0,\Delta\to-\infty\quad h_x=h_y=0$,  The state is given by a path
\bea
\ket{0}_B=(\ldots,-,+,-),\qquad \ket{1}_B=(\ldots,+,-,+),
\ena
and nonzero $h_z$ resoloves the degeneration of the ground states. If we apply small $h_x,h_y$, say $\delta\times h_x$, then we get for $\H^0_B$
\bea
\ket{0}_B=(\ldots,-,+,-)-\delta (\ldots,-,+,+)
\ena 
and there is a finite gap between the next lowest energy state $(\ldots,-,+,+)+\delta (\ldots,-,+,-)$.
The case of the XXZ model with a diagonal $K$-matrix is discussed with $q$-expansion in \cite{bdry XXZ}.



\setcounter{section}{2}
\setcounter{equation}{0}
\section{The boundary SOS model}
\lb{sec:SOS}
With some modifications, we recall the result of \cite{MW} which concerns the boundary problem for the ABF model with a diagonal $K$-matirx. In \cite{MW}, by following \cite{LuP} and \cite{bdry XXZ}, the boundary transfer matrix $T^{(k)}_B(u)$ is bosonised and the eigenstates of $T^{(k)}_B(u)$ are constructed in Fock spaces. 

\subsection{The SOS model}
Local Boltzmann weight of bulk type is given as follows \cite{ABF}, we follow the convention of \cite{LaP} (Fig.\ref{fig:W-K}).
\bea
&&
W\left(
\begin{array}{cc}
a & a\pm 1 \\  a\pm 1 & a\pm 2
\end{array} \Biggr| u   \right) =R_0(u),\\
&&
W\left(
\begin{array}{cc}
a & a\pm 1 \\  a\pm 1 & a 
\end{array} \Biggr| u   \right) =R_0(u){[1][a\pm u]\over [a][1-u]} ,\\
&&
W\left(
\begin{array}{cc}
a & a\pm 1 \\  a\mp 1 & a
\end{array} \Biggr| u   \right) =R_0(u){[a\pm 1][u]\over [a][1-u]},\\[2mm]
&&
[u]=x^{{u^2\over r}-u} \Theta_{x^{2r}}(x^{2u}),\quad\Theta_p(z)=(z;p)_\infty(pz^{-1};p)_\infty(p;p)_\infty,
\ena
$R_0(u)$ is given in \eqref{eqn:norm-R}.
The site variable $a$ called height is an integer and satisfies the admissiblity: for variables $a_1,a_2$ at adjacent sites
\bea
|a_1-a_2|=1.
\ena
The difference between our formulation and \cite{MW} is that we do not impose the restriction
\bea
1\leq a\leq r-1.
\ena

We consider the so called regime III:
\bea
\eps>0,\quad r\geq 1,\quad 0<u<1.
\lb{eqn:regIII}
\ena

The Boltzmann weight satisfies the relations\\
\noi
(i) Star-triangle relation
\bea
&&
\sum_g W\BW fgab{u-v}
W\BW gdbcu 
W\BW fegdv \non\\
&&
\quad = 
\sum_g W\BW agbcv 
W\BW feagu 
W\BW edgc{u-v} ,
\label{star-tri}
\ena
(ii) Unitarity relation
\bea
\sum_g W\BW dgabu 
W\BW dcgb{-u} 
=\delta_{ac} ,
\label{uni-ABF}
\ena
(iii) Crossing relation
\bea
W\BW dcabu 
=\frac{[a]}{[b]}
W\BW adbc{1-u} .
\label{cr}
\ena



Lattice vertex operators of face type (half infinite transfer matrices) $\phi(u)^{n'}_{n},\,\phi^*(u)^{n'}_{n},\,n'=n\pm1$ are defined as in the Fig.\ref{fig:VO-f} \cite{LuP}. A face-path is a semi-infinite sequence of integers $(\ldots,a_2,a_1)$. Among them, a $(l,k)$-path is the one having central heights $k$ and the boundary heights $(l,l+1)$, that is $(\ldots,l+1,l,l+1,l,\ldots,k)$. We denote by $\H_{l,k}$ the eigenspace of the corner transfer matrix \cite{ABF} in the NW quadrant spanned by $(l,k)$-paths. Then
\bea
\phi(u)_k^{k+\ve},\phi^*(u)_k^{k+\ve}:\,\H_{l,k}\to\H_{l,k+\ve},
\ena
From graphical argument, following relations can be derived\\
\noi
(i) Commutation relation
\bea
&&
\phi(u_2)^a_b \phi(u_1)^b_c  =
\sum_{g} W\BW agbc{u_2-u_1}\phi(u_1)^a_g \phi(u_2)^g_c,
\lb{eqn:VO-comm}
\ena
(ii) Duality
\bea
\phi^*(u)_k^{k+\ve}=\frac{1}{[k]} \phi(1-u)_k^{k+\ve},
\lb{eqn:VO-dual}
\ena
(iii) Inversion relation
\bea
\sum_g \phi^{*}(u)^a_g \phi(u)^g_a = 1, \quad
       \phi(u)^a_b \phi^{*}(u)^b_c = \delta_{ac}.
\lb{eqn:VO-inv}
\ena


\subsection{The boundary SOS model}


The boundary weight: $K$-matrix is given as a solution of the eqaution:\\
\noi
(iv) Reflection equation
\bea
&&
\sum_{f,g}
W\BW cfba{u-v} W\BW cdfg{u+v} K\Kmatrix fgau K\Kmatrix degv \\
&&=\quad \sum_{f,g}
W\BW cdfe{u-v} W\BW cfbg{u+v} K\Kmatrix fegu K\Kmatrix bgav.
\lb{eqn:RE}
\ena
In \cite{MW}, the diagonal solution found by \cite{BPO} is used (Fig.\ref{fig:W-K})
\bea
\frac{K\Kmatrix{k+1}{k}{k}{u}}{K\Kmatrix{k-1}{k}{k}{u}}=\frac{[c+u][k+c-u]}{[c-u][k+c+u]}.
\ena
There are two regions A and B depending on the parameter $c$
\bea
&&
\mbox{region A}:\:\:x^{2c}=-x^{2b},\;-1<b<1,\non\\
&&
\mbox{region B}:\:\:x^{2c}=x^{2b},\;-1<b<1,
\lb{eqn:b}
\ena
and $u$ is restricted to satisfy
\bea
0<u<|b|<1.
\ena

Then the unnormalized boundary transfer matrix is defined as
\bea
T^{(k)}_B(u)=\sum_{\ve=\pm 1} \phi^*(-u)^k_{k+\ve} K\Kmatrix{k+\ve}{k}{k}{u} \phi(u)^{k+\ve}_k,
\ena
and we will denote its eigenspace spanned by $(l,k)$-paths as $\H^B_{l,k}$.
The reflection equation \eqref{eqn:RE} implies the integrability of the model
\be
[T^{(k)}_B(u),T^{(k)}_B(v)]=0.
\en

The ground state of the model depends on $c$ as
\bea
\frac{K\Kmatrix{k+1}{k}{k}{u}}{K\Kmatrix{k-1}{k}{k}{u}}=
\left\{
 \begin{array}{rl}
  & >1\quad \mbox{if}\quad b>0,\\
  & <1\quad \mbox{if}\quad b<0.
 \end{array}\right.
\ena
That is the ground state belongs to $\H^B_{k,k},\,\H^B_{k-1,k}$ for $b>0,\,b<0$ respectively.

The normalizaiton of the $K$-matrix is so chosen that the largest eigenvalue of the boundary transfer matrix $T^{(k)}_B(u)$ is $1$. It also depends on the positivity of $b$. For $b>0$ it is given by
\bea
&&
K^{(c)}_{>}\Kmatrix {k+1}kku = h^{(k)}_>(u),\lb{eqn:norm-K>+}\\
&&
K^{(c)}_{>}\Kmatrix {k-1}kku = h^{(k)}_>(u)
{[c-u][k+c+u]\over [c+u][k+c-u]},\lb{eqn:norm-K>-}\\[2mm]
&&
h^{(k)}_>(u) = \zeta^{r-1-2k\over 2r}
{f(\zeta)p^{(k)}_>(\zeta)p^{(k)}_>(x^2\zeta^{-1})\over
f(\zeta^{-1})p^{(k)}_>(\zeta^{-1})p^{(k)}_>(x^2\zeta)},\\[2mm]
&&
f(\zeta) =
{(x^{2r}\zeta^2;x^8,x^{2r})_\infty
(x^{8}\zeta^2;x^8,x^{2r})_\infty\over
(x^{6}\zeta^2;x^8,x^{2r})_\infty
(x^{2+2r}\zeta^2;x^8,x^{2r})_\infty},\\[2mm]
&&
p^{(k)}_>(\zeta) =
{
\{x^{2(1+c)}\zeta\}\{x^{2(r-c-k+1)}\zeta\}
\over
\{x^{2(r-c)}\zeta\}\{x^{2(c+k)}\zeta\}
},
\ena
and for $b<0$ 
\bea
&&
K^{(c)}_<\Kmatrix{k+1}{k}{k}{u}=h_<^{(k)}(u){\frac{[c+u][k+c-u]}{[c-u][k+c+u]}},\label{eqn:norm-K<+}\\
&&
K^{(c)}_<\Kmatrix{k-1}{k}{k}{u}=h_<^{(k)}(u),
\label{eqn:norm-K<-}\\[2mm]
&&
h_<^{(k)}(u)=
\zeta^{\frac{2k-1-r}{2r}}
{\frac{f(\zeta)p_<^{(k)}(\zeta)p_<^{(k)}(x^2\zeta^{-1})}
{f(\zeta^{-1})p_<^{(k)}(\zeta^{-1})p_<^{(k)}(x^2\zeta)}},\\[2mm]
&&
p_<^{(k)}(\zeta)=
\frac{
\{x^{2(1-c)}\zeta\}\{x^{2(c+k+1)}\z\}
}{
\{x^{2(r+c)}\zeta\}\{x^{2(r-c-k)}\zeta\}
}.
\ena
There are two more relations satisfied by $K$-matrix\\
\noi
(v) Boundary unitarity relation
\bea
K^{(c)}_{>\atop <}\Kmatrix{k'}{k}{k}{u}K^{(c)}_{>\atop <}\Kmatrix{k'}{k}{k}{-u}=1,
\ena
(vi) Boundary crossing relation
\bea
K^{(c)}_{>\atop <}\Kmatrix{k'}{k}{k}{1-u}=\sum_{k''=k\pm 1}\frac{[k'']}{[k]}W\BW {k'}kk{k''}{-2u+2} K^{(c)}_{>\atop <}\Kmatrix{k''}{k}{k}{u}.
\ena


For definiteness, we write down the boundary transfer matrix in the normalised form
\bea
T^{(k)}_B(u)=\sum_{\ve=\pm 1} \phi^*(-u)^k_{k+\ve} K^{(c)}_{>\atop <}\Kmatrix{k+\ve}{k}{k}{u} \phi(u)^{k+\ve}_k.
\ena

\footnote[0]{Eqn.(2.17) of \cite{MW} is misprinted and arguments of ${\bar T}^{(k)}_B$ should be exchanged.}


\subsection{Bosonization}
\lb{subsec:boson}

Bosonization of the objects introduced in the previous sections are achieved with the following oscillators
 \cite{LuP}.
\bea
&&
[ \beta_n,\beta_m]=\delta_{n+m,0}
{[n]_x^3\over n[2n]_x}{[\rs n]_x\over [rn]_x},\\
&&
\alpha_n=\frac{[2n]_x}{[n]_x}\beta_n,\\
&&
[P,iQ]=1,\\
&&
[n]_x={x^n-x^{-n}\over x-x^{-1}},\\
&&
\rs=r-1,
\ena
and the Fock space ${\cal F}_{l,k}$ and its dual ${\cal F}^*_{l,k}$ are defined as
\bea
&&
{\cal F}_{l,k}={\mathbb C}[\beta_{-1},\beta_{-2},\cdots]|l,k\rangle,\\
&&
\beta_n\ket{l,k}=0,\quad\mbox{for}\quad n>0,\\
&&
P |l,k\rangle=
\left(-l\sqrt{r\over 2\rs}+k\sqrt{\rs \over 2r} \right)
|l,k\rangle,\\[2mm]
&&
{\cal F}^*_{l,k}=\bra{l,k}{\mathbb C}[\beta_{1},\beta_{2},\cdots],\\[2mm]
&&
\bra{l,k}\beta_n=0,\quad\mbox{for}\quad n<0,\\
&&
\bra{l,k}P =
\bra{l,k}
\left(-l\sqrt{r\over 2\rs}+k\sqrt{\rs \over 2r} \right),\\
&&
\br{l,k|l,k}=1.
\ena
For $\H^B_{l,k}$ we assume
\bea
\H^B_{l,k}\subset\H_{l,k}\cong\F_{l,k}.
\ena

Bosonic vertex operators are given as
\bea
&&
\bPhi_\ve(u)=\Phi_\ve(u){\ve^K \over [K]},\\
&&
\Phi_-(u)= 
:\exp\left(- \sum_{n\neq 0} {\beta_n\over [n]_x}\z^{-n}\right): \times
e^{\sqrt{  \rs\over 2  r}iQ}
\z^{\sqrt{  \rs\over 2  r}P+{ \rs\over 4 r}},\\
&&
\Phi_+(u)=
\oint_C {d\z_1\over 2\pi i\z_1} \Phi_-(u) x_-(u_1) 
{[u-u_1-1/2+K] \over [u-u_1+1/2]},\\
&&
x_-(u)= 
:\exp\left( \sum_{n\neq 0} {\alpha_n\over [n]_x}\z^{-n}\right): \times
e^{ -\sqrt{2 \rs\over  r}iQ}
\z^{ -\sqrt{2 \rs\over  r}P+{ \rs\over r}},\\
&&
K\Bigr|_{ {\cal F}_{l,k}}= k \times {\rm id}_{ {\cal F}_{l,k}},\\
&& 
x_-(u):\F_{l,k}\to \F_{l,k-2},\\
&&
\Phi_-(u):\F_{l,k}\to \F_{l,k+1},\\
&&
\Phi_+(u):\F_{l,k}\to \F_{l,k-1}.
\ena
$x_-(u)$ is a screening current.

Commutation relation, duality and inversion relation for $\bPhi_{\ve}(u)$ follow.
\bea
&&
\bPhi_{\ve_2}(u_2)\bPhi_{\ve_1}(u_1)=\sum_{{\ve_1',\ve_2'}\atop{\ve_1'+\ve_2'=\ve_1+\ve_2}}
W\left(
\begin{array}{cc}
K & K+\ve_1' \\  K+\ve_2 & K+\ve_1+\ve_2 
\end{array} \Biggr| u_2-u_1
  \right)
\bPhi_{\ve_1'}(u_1)\bPhi_{\ve_2'}(u_2),\non\\
&&\\
&&
\bPhi_\ve^*(u)=\ve\bPhi_{-\ve}(u-1)[K](-)^K,\\
&&
g\;\sum_{\ve=\pm}\bPhi^*_{\ve}(u)\bPhi_\ve(u)=1,\\
&&
g\;\bPhi_\ve(u)\bPhi^*_{\ve'}(u)=\delta_{\ve\ve'},\\
&&
g=x^{-\frac{r^*}{2r}}(x^2)_\infty(x^{2r})_\infty\frac{\{x^2\}\{x^{2r+2}\}}{\{x^4\}\{x^{2r+4}\}}.
\ena
Hence we identify lattice vertex operators and bosonic ones as
\bea
&&
\phi(u)^{k}_{k+\ve}=\bPhi_\ve(-u),\\
&&
\phi^*(u)^{k+\ve}_{k}=\bPhi^*_\ve(-u).
\ena
Then the boundary transfer matrix in the bosonic language is
\bea
T^{(k)}_B(u)=g\;\sum_{\ve=\pm} \bPhi^*_\ve(u) K^{(c)}_{>\atop <}\Kmatrix{k-\ve}{k}{k}{u} \bPhi_\ve(-u).
\ena

\noi
{\em Remark}\quad $\bPhi_\ve(u),\bPhi^*_\ve(u)$ serve as vertex operators for \cite{LaP} where (3.15) of \cite{LaP} is modified as $g\sum_{n'}(n-n')[n'](-)^{n+1}\Phi(u-1)^n_{n'}\Phi(u)_n^{n'}=1$ and fixing the parameter $m=1$.


\subsection{Boundary vacuum states}
\lb{sec:bvs}
The maximal eigenvectors of $T^{(k)}_B(u)$ are called the boundary vacuum vectors. Under the normalization of the $K$-matrix \eqref{eqn:norm-K>+}, \eqref{eqn:norm-K>-}, \eqref{eqn:norm-K<+}, \eqref{eqn:norm-K<-} they satisfy  
\bea 
&&
T_B^{(k)}(u) \rbs k{k} = \rbs k{k}\in \F_{k,k}, \quad\quad 
\mbox{for}\quad b>0,\label{cond0}\\[2mm]
&&
T_B^{(k)}(u) \rbs {k-1}k = \rbs {k-1}k\in \F_{k-1,k} , \quad\quad 
\mbox{for}\quad b<0,\label{cond1}\\[2mm]
&&
\lbs{k}{k}T_B^{(k)}(u) = \lbs{k}{k} \in \F_{k,k}^{*}
\quad\quad 
\mbox{for}\quad b>0,\label{dcond0}\\[2mm]
&&
 \lbs{k-1}{k}T_B^{(k)}(u) = \lbs{k-1}{k} \in \F_{k-1,k}^{*}\
\quad\quad 
\mbox{for}\quad b<0,\label{dcond1}
\ena
It is easily shown with \eqref{eqn:VO-inv} that these are equivalent to
\bea
&&
K^{(c)}_{>\atop <}\Kmatrix{k-\ve}{k}{k}{u}\bPhi_\ve(-u)|k-i,k\rangle^c_B=\bPhi_\ve(u)|k-i,k\rangle^c_B,\\
&&
{}^c_B\langle k-i,k|\bPhi^*_\ve(u)K^{(c)}_{>\atop <}\Kmatrix{k-\ve}{k}{k}{u}={}^c_B\langle k-i,k|\bPhi^*_\ve(-u).
\lb{eqn:cond-bvs}
\ena 
Their explicit form is obtained by solving these equations \cite{MW}
\bea
&&
|k-i,k\rangle^c_B=e^{F_i^{c,k}}|k-i,k\rangle,\\[2mm]
&&
F_i^{c,k}= -\frac12 \sum_{m>0}  \ka_m\beta_{-m}{}^2 +
\sum_{m>0} \beta_{-m}D_{m,i}^{c,k} ,\\[2mm]
&&
e^{-F_i^{c,k}} \beta_{m} e^{F_i^{c,k}} = \beta_m - \beta_{-m} + 
{D_{m,i}^{c,k}\over \ka_m}, \\[2mm]
&&
e^{-F_i^{c,k}} \beta_{-m} e^{F_i^{c,k}} = \beta_{-m},\\[2mm]
&&
D_{m,0}^{c,k} =
- {[(k-1)m]_x[(r-2c-k)m]^+_x \over [m]_x[\rs m]_x} 
- \theta_m\!\left( 
{[m/2]_x[rm/2]^+_x \over [m]_x[\rs m/2]_x}
\right),\label{D0def}\\[3mm]
&&
D_{m,1}^{c,k} = D_{m,0}^{c,k}|_{{k\to k-r} \atop{c\to c+r}},
\ena
and
\bea
&&
{}^c_B\langle k-i,k|=\langle k-i,k|e^{G_i^{c,k}},\\[2mm]
&&
G_i^{c,k}= -\frac12 \sum_{m>0} x^{4m}\ka_m\beta_{m}{}^2 +
\sum_{m>0} \beta_{m}E_{m,i}^{c,k} ,\\[2mm]
&&
e^{-G_i^{c,k}} \beta_{-m} e^{G_i^{c,k}} = \beta_{-m} - x^{4m}\beta_{m} + 
{E_{m,i}^{c,k}\over\ka_m }, \\[2mm]
&&
e^{-G_i^{c,k}} \beta_{m} e^{G_i^{c,k}} = \beta_{m},\\[2mm]
&&
E_{m,1}^{c,k} = 
x^{2m} { [(r-k-1)m]_x[(2c+k)m]^+_x \over [m]_x[\rs m]_x}
+x^{2m} \theta_m\!\left(
{[m/2]_x[rm/2]^+_x \over [m]_x[\rs m/2]_x} \right),\\[2mm]
&&
E_{m,0}^{c,k} = E_{m,1}^{c,k}|_{{k\to k+r} \atop{c\to c-r}},
\ena
where
\bea
&&
\theta_m=\cases{x& if $m$ is even;\cr
0& otherwise.\cr},\\[2mm]
&&
[m]^+_x=x^k+x^{-k}.
\ena
Note that as a linear combination of paths, the boundary vacuum states coincides for the restricted SOS and unrestricted SOS model.

\subsection{Alternative bosonization}
\lb{subsec:another-boson}
In Sec.\ref{sec:corr-fn}, we need another bosonization such that
\bea
\H^B_{l,k}\subset\H_{l,k}\cong\F_{-l,-k}.
\ena
It is given as follows. Vertex operators are realized as
\bea
&&
\phi(u)^{k+\ve}_k=\bPhi_\ve(-u),\\
&&
\phi^*(u)_{k+\ve}^k=\bPhi^*_\ve(-u).
\ena
As for the boundary vacuum states, conditions \eqref{eqn:cond-bvs} are changed as
\bea
&&
K^{(c)}_{>\atop <}\Kmatrix{k+\ve}{k}{k}{u}\bPhi_\ve(-u)|\overline{k-i,k}\rangle^c_B=\bPhi_\ve(u)|\overline{k-i,k}\rangle^c_B,\\
&&
{}^c_B\langle \overline{k-i,k}|\bPhi^*_\ve(u)K^{(c)}_{>\atop <}\Kmatrix{k+\ve}{k}{k}{u}={}^c_B\langle \overline{k-i,k}|\bPhi^*_\ve(-u),
\ena
and the states are given by
\bea
&&
|\overline{k-i,k}\rangle^c_B=e^{{\bar F}^{c,k}_i}|-k+i,-k\rangle,\\[2mm]
&&
{\bar F}^{c,k}_i=F^{c,k}_i|_{D^{c,k}_{n,i}\to{\bar D}^{c,k}_{n,i}}\\
&&
{\bar D}^{c,k}_{n,i}=D^{c,k}_{n,1-i}|_{{c\to -c}\atop{k\to r-k}},\\
&&
{}^c_B\langle \overline{k-i,k}|=\langle -k+i,-k|e^{{\bar G}^{c,k}_i},\\[2mm]
&&
{\bar G}^{c,k}_i=G^{c,k}_i|_{E^{c,k}_{n,i}\to{\bar E}^{c,k}_{n,i}},\\
&&
{\bar E}^{c,k}_{n,i}=E^{c,k}_{n,1-i}|_{{c\to -c}\atop{k\to r-k}}.
\ena


\setcounter{section}{3}
\setcounter{equation}{0}
\section{Correlation functions of the boundary XYZ model}
\lb{sec:corr-fn}
In this section, we consider the correlation functions for the sector $i=0$ of $\H^i_B$. The case $i=1$ can be handled in the same way.

\subsection{Face-vertex correspondence of the bulk weights}
We recall the face-vertex correspondence of the bulk weights \cite{Bax76}, \cite{LaP}. The intertwining vectors are given as
\footnote[1]{We fix the parameter $m=1$ in \cite{LaP}.}
\bea
&&
t_+(u)_n^{n'}=\frac{1}{\sqrt{2}}\;\vartheta_3\!\left({(n'-n)u+n'\over 2r};{i\pi\over 2\epsilon r}\right),\\
&&
t_-(u)_n^{n'}=\frac{(-)^n}{\sqrt{2}}\;\vartheta_0\!\left({(n'-n)u+n'\over 2r};{i\pi\over 2\epsilon r}\right),
\ena
where $n'=n\pm 1$.
Its conjugate: $t_\ve^*(u)_n^{n'}$, primed: $t'_\ve(u)_n^{n'}$ and primed conjugate: $t_\ve'^*(u)^{n'}_n$ are given by (Fig.\ref{fig:vector})
\bea
&&
t_\ve^*(u)_n^{n'}=(-)^nC^2{n'-n\over [n][u]}t_{-\ve}(u-1)_n^{n'},\\
&&
t'_\ve(u)_n^{n'}
={[u]\over[u-1]}{[n']\over[n]}\,t_\ve(u-2)_n^{n'},\\
&&
t_\ve'^*(u)^{n'}_n=\frac{[u-1][n]}{[u][n']}t_\ve^*(u-2)^{n'}_n,\\
&&
C=\sqrt{{\pi\over\epsilon r}}e^{\epsilon r/4}.
\lb{eqn:C}
\ena
They satisfy the following relations (Fig.\ref{fig:vec-rel})
\bea
&& 
\sum_{\ve=\pm} t^*_\ve(u)^{n'}_n t_\ve(u)_{n''}^n = \delta_{n'n''},
\qquad
\sum_{s=n\pm 1}t_{\ve'}^*(u)^{s}_n t_\ve(u)^{n}_s=\delta_{\ve'\ve},\non\\
&&
\sum_{\ve=\pm} t^*_\ve(u)^n_{n'} t'_\ve(u)_n^{n''} = \delta_{n'n''},
\qquad
\sum_{s=n\pm 1}t_{\ve'}^*(u)_{s}^n t'_\ve(u)_{n}^s=\delta_{\ve'\ve},\non\\
&&
\sum_{\ve=\pm} t_{\ve}'^*(u)^{n'}_nt_\ve'(u)^{n}_{n}=\delta_{n'n''} ,
\qquad
\sum_{s=n\pm 1}t_{\ve'}'^*(u)^{s}_n t'_\ve(u)^{n}_s=\delta_{\ve'\ve}.
\lb{eqn:vec-rel}
\ena
The basic face-vertex correspondence is (Fig.\ref{fig:fv-RW})
\bea
\sum_{\ve'_1,\ve'_2=\pm}R(u-v)_{\ve_1\ve_2}^{\ve'_1\ve'_2}
t_{\ve'_1}(u_0-u)^{n'}_{s'} t_{\ve'_2}(u_0-v)^{s'}_n
=\sum_{ s\in\mathbb{Z} } t_{\ve_2}(u_0-v)^{n'}_s t_{\ve_1}(u_0-u)^s_n
W\BW {n'}{s'}sn{u-v},\non\\
&&
\lb{eqn:fv-RW}
\ena
and from this identity and \eqref{eqn:vec-rel} we get the variants, for instance
\bea
\sum_{\ve_1',\ve_2'=\pm}
t^*_{\ve_2'}(u_0-v)_{n'}^s t^*_{\ve_1'}(u_0-u)_s^n
R(u-v)_{\ve'_1\ve'_2}^{\ve_1\ve_2}
=\sum_{s'\in\mathbb{Z}}
W\BW {n'}{s'}sn{u-v}
t^*_{\ve_1}(u_0-u)_{n'}^{s'} t^*_{\ve_2}(u_0-v)_{s'}^n,
&&\non\\
&&\\
\sum_{\ve'_1,\ve'_2=\pm}
t'^*_{\ve'_2}(u_0-v)_{s}^{n'} R(u-v)^{\ve'_1\ve_2}_{\ve_1\ve'_2}
t'_{\ve'_1}(u_0-u)_{n}^{s} 
=\sum_{ s'\in\mathbb{Z} } t'_{\ve_1}(u_0-u)^{n'}_{s'}
W\BW {s}n{n'}{s'}{u-v} t'^*_{\ve_2}(u_0-v)^{s'}_n .
&&\non\\
&&
\ena
Note that the principal regime of the XYZ model \eqref{eqn:pr-reg} is mapped to the regime III of the SOS model \eqref{eqn:regIII}.



\subsection{Face-vertex correspondence of $K$-matrix}
\lb{subsec:fv-K}
We make a $K$-matrix of vertex type $K(u;c,l,u_0)$ from the $K$-matrix of Sec.\ref{sec:SOS} as
\bea
&&
K(u;c,l,u_0)^\ve_{\ve'}=\sum_{\nu=\pm 1}t_{\ve}^*(u_0-u)^{l}_{l+\nu} t_{\ve'}'(u_0+u)^{l+\nu}_l K^{(c)}_<\Kmatrix{l+\nu}{l}{l}{u}.
\lb{eqn:fv-K}
\ena
This type of face-vertex correspondence was found in \cite{FHLS}.
It can be easily verified by graphical argument that this $K$-matrix satisfies the boundary Yang-Baxter equation \eqref{eqn:bdryYBE}. As we will see below, through this face-vertex correspondence of $K$-matrix and those of $R$-matrix, the sector $i=0$ of the boundary XYZ model is mapped to the region $b<0$ of the boundary SOS model. And this fixes the normalization of the $K$-matrix of the SOS model in \eqref{eqn:fv-K}. 

Explicit form of elements are
\bea
&&
K(u;c,l,u_0)^\ve_{\ve'}=\frac{C^2[u_0+u]}{2[u_0-u][u_0+u-1][l]}\frac{h^{(l)}_<(u)}{[c-u][l+c+u]}{\bar K}(u;c,l,u_0)^\ve_{\ve'},
\lb{eqn:K-bar}\\[2mm]
&&
{\bar K}(u;c,l,u_0)^+_{+}=
-\vartheta_0\!\left({u+\delta+l\over 2r};{i\pi\over 2\epsilon r}\right)\vartheta_3\!\left({u-\delta+l\over 2r};{i\pi\over 2\epsilon r}\right)
[c+u][l+c-u]\non\\[2mm]
&&
\hspace{3cm} +\vartheta_0\!\left({u+\delta-l\over 2r};{i\pi\over 2\epsilon r}\right)\vartheta_3\!\left({u-\delta-l\over 2r};{i\pi\over 2\epsilon r}\right)[c-u][l+c+u],\non\\[2mm]
&&
{\bar K}(u;c,l,u_0)^-_{-}=\vartheta_3\!\left({u+\delta+l\over 2r};{i\pi\over 2\epsilon r}\right)\vartheta_0\!\left({u-\delta+l\over 2r};{i\pi\over 2\epsilon r}\right)[c+u][l+c-u]\non\\[2mm]
&&
\hspace{3cm} -\vartheta_3\!\left({u+\delta-l\over 2r};{i\pi\over 2\epsilon r}\right)\vartheta_0\!\left({u-\delta-l\over 2r};{i\pi\over 2\epsilon r}\right)[c-u][l+c+u],\non\\[2mm]
&&
{\bar K}(u;c,l,u_0)^+_{-}=(-)^{l+1}\vartheta_0\!\left({u+\delta+l\over 2r};{i\pi\over 2\epsilon r}\right)\vartheta_0\!\left({u-\delta+l\over 2r};{i\pi\over 2\epsilon r}\right)[c+u][l+c-u]\non\\[2mm]
&&
\hspace{3cm} +(-)^l\vartheta_0\!\left({u+\delta-l\over 2r};{i\pi\over 2\epsilon r}\right)\vartheta_0\!\left({u-\delta-l\over 2r};{i\pi\over 2\epsilon r}\right)[c-u][l+c+u],\non\\[2mm]
&&
{\bar K}(u;c,l,u_0)^-_{+}=(-)^l\vartheta_3\!\left({u+\delta+l\over 2r};{i\pi\over 2\epsilon r}\right)\vartheta_3\!\left({u-\delta+l\over 2r};{i\pi\over 2\epsilon r}\right)[c+u][l+c-u]\non\\[2mm]
&&
\hspace{3cm} +(-)^{l+1}\vartheta_3\!\left({u+\delta-l\over 2r};{i\pi\over 2\epsilon r}\right)\vartheta_3\!\left({u-\delta-l\over 2r};{i\pi\over 2\epsilon r}\right)[c-u][l+c+u],\non\\[2mm]
&&\\
&&
\delta=-u_0+1.
\ena

We show below that the diagonal $K$-matrix of \cite{bdry qKZ} is a sepcial case of this $K(u;c,l,u_0)$. Our convention of parameters are that of \cite{MW}, \cite{LaP} and the correspondence with \cite{bdry qKZ} is as follows. We attach the subscript \lq\lq $d$" to those of \cite{bdry qKZ}.
\bea
&&
x=-q_d,\quad \frac{\epsilon}{\pi}=\frac{\lambda_d}{2K_d},\non\\
&&
x^{2r}=p_d,\quad\mbox{(elliptic nome)}\non\\[2mm]
&&
k=k_d,\quad\mbox{(elliptic modulus)}\non\\[2mm]
&&
K=K_d,\quad\mbox{(half-period magnitude)}\non\\[2mm]
&&
\zeta=\z_d^{-2},\quad \frac{\epsilon u}{\pi}=\frac{u_d}{2K_d},\non\\[2mm]
&&
R(u)|_{d(u)\to -d(u)}=R_d(\z_d),
\ena
Then the diagonal $K$-matrix is as follows. 
\bea
&&
K(\zeta_d)=\frac{1}{f(\zeta_d;r)}{\widehat K_d}(\zeta_d;r_d),\quad \z_d=e^{\frac{\pi u_d}{2K_d}}\\[2mm]
&&
{\widehat K}(\zeta_d;r_d)=
\left(\frac{\mbox{snh}(\eta_d+u_d)}{\mbox{snh}(\eta_d-u_d)}\quad\atop\qquad\qquad\quad 1\right),\quad r_d=e^{\sstyle{\frac{\pi\eta_d}{K_d}}},\mbox{ for } 0<r_d<1,\\[2mm]
&&
{\widehat K}(\zeta_d;r_d)=
\left(\frac{\mbox{snh}(\eta_d+iK_d+u_d)}{\mbox{snh}(\eta_d+iK_d-u_d)}\quad\atop\qquad\qquad\quad 1\right),\quad r_d=-e^{\sstyle{\frac{\pi\eta_d}{K_d}}},\mbox{ for } -1<r_d<0.
\ena
The parameter $r_d$ is for the magnetic field $h_z$ and $-1<r_d<1$ corresponds to the sector $i=0$. The parameter $\eta_d$ satisfies
\bea
0<u_d<-\eta_d<\lambda_d,\;\Leftrightarrow\;0<u<-\frac{\pi}{2K_d\epsilon}\eta_d<1.\lb{eqn:eta}
\ena

Set
\bea
\frac{\delta}{r}=\frac 12+{i\pi\over 2\epsilon r},\quad l=\frac r2-c,
\lb{eqn:cond-diag}
\ena
then off diagonal elements vanish and diagonal elements are
\bea
&&
{\bar K}(u;c,l,u_0)^+_{+}=
-C^2e^{\frac{\pi^2}{4\epsilon r}}
\frac{
\vartheta_1\!\left(\frac 12;{i\pi\over \epsilon r}\right)
\vartheta_2\!\left(0;{i\pi\over 2\epsilon r}\right)
}{
\vartheta_1^2\!\left(\frac 14;{i\pi\over 2\epsilon r}\right)
}
\non\\[2mm]
&&
\hspace{3cm}\times
\vartheta_2\!\left(\frac ur;{i\pi\over 2\epsilon r}\right)\vartheta_2\!\left(\frac cr; {i\pi\over \epsilon r}\right)\vartheta_2\!\left(\frac{u-c}{2r};{i\pi\over 2\epsilon r}\right)\vartheta_1\!\left(\frac{u+c}{2r};{i\pi\over 2\epsilon r}\right) ,
\\[2mm]
&&
{\bar K}(u;c,l,u_0)^-_{-}=C^2e^{\frac{\pi^2}{4\epsilon r}}\frac{\vartheta_1\!\left(\frac 12;{i\pi\over \epsilon r}\right)\vartheta_2\!\left(0;{i\pi\over 2\epsilon r}\right)}{\vartheta_1^2\!\left(\frac 14;{i\pi\over 2\epsilon r}\right)}\non\\[2mm]
&&
\hspace{3cm}\times\vartheta_2\!\left(\frac ur;{i\pi\over 2\epsilon r}\right)\vartheta_2\!\left(\frac cr; {i\pi\over \epsilon r}\right)\vartheta_1\!\left(\frac{u-c}{2r};{i\pi\over 2\epsilon r}\right)\vartheta_2\!\left(\frac{u+c}{2r};{i\pi\over 2\epsilon r}\right) .
\ena
From this we have
\bea
&&
\frac{{\bar K}(u;c,l,u_0)^+_{+}}{{\bar K}(u;c,l,u_0)^-_{-}}=-\frac{\vartheta_2\!\left(\frac{u-c}{2r};{i\pi\over 2\epsilon r}\right)\vartheta_1\!\left(\frac{u+c}{2r};{i\pi\over 2\epsilon r}\right)}{\vartheta_1\!\left(\frac{u-c}{2r};{i\pi\over 2\epsilon r}\right)\vartheta_2\!\left(\frac{u+c}{2r};{i\pi\over 2\epsilon r}\right)}\non\\[2mm]
&&
\hspace{3cm}
=-\frac{\vartheta_0\!\left(\frac{\epsilon}{i\pi}(u-c);{i2\epsilon r\over \pi}\right)\vartheta_1\!\left(\frac{\epsilon}{i\pi}(u+c);{i2\epsilon r\over \pi}\right)}{\vartheta_1\!\left(\frac{\epsilon}{i\pi}(u-c);{i2\epsilon r\over \pi}\right)\vartheta_0\!\left(\frac{\epsilon}{i\pi}(u+c);{i2\epsilon r\over \pi}\right)}\non\\[2mm]
&&
\hspace{3cm}
=-\frac{\mbox{snh}\!\left(u_d+\frac{2K\epsilon}{\pi}c;k_d \right)}{\mbox{snh}\!\left(u_d-\frac{2K\epsilon}{\pi}c;k_d \right)}.
\ena
Therfore the parameter for the magnetic filed $\eta_d$ is identified with $c$ as
\bea
&&
\eta_d+iK_d=\frac{2K_d\epsilon}{\pi}c\quad\mbox{for}\quad -1<r_d<0,\\
&&
\eta_d=\frac{2K_d\epsilon}{\pi}c\quad\mbox{for}\quad 0<r_d<1.
\ena
Thus, from \eqref{eqn:eta}, the region considered in \cite{bdry qKZ} is mapped to the regions of the SOS model \eqref{eqn:b} as
\bea
&&
-1<r_d<0\:\to\:
0<u<-c+\frac{i\pi}{2\epsilon}<1,\:(\:\mbox{i.e.}\:b<0\:\;\mbox{of region A}),\label{eqn:regA}\\[2mm]
&&
0<r_d<1\:\to\:
0<u<-b<1,\:(\:\mbox{i.e.}\:b<0\:\;\mbox{of region B}),\label{eqn:regB}
\ena

$K(u;c,l,u_0)$ does not satsify the boundary crossing relation and $K_d(\zeta_d;r_d)$ and $K(u;c,l={r\over 2}-c,u_0=1-{r\over 2}-{i\pi\over 2\epsilon})$ differ by over all factor. Therefore we should change the intertwining vectors as
\bea
&&
t^{new}_\ve(u)^{n'}_n=f(u)t_\ve(u)^{n'}_n,\non\\
&&
t^{*\,new}_\ve(u)^{n'}_n=\frac{1}{f(u)}t^*_\ve(u)^{n'}_n,\non\\
&&
t'^{new}_\ve(u)^{n'}_n=f(u)t'_\ve(u)^{n'}_n,\non\\
&&
t^{*\,new}_\ve(u)^{n'}_n=\frac{1}{f(u)}t^*_\ve(u)^{n'}_n,
\ena
where $f(u)$ satisfies
\bea
&&
\frac{f(u_0-1+u)}{f(u_0+1-u)}\frac{f(u_0+u)}{f(u_0-u)}=\frac{[u_0-u+1]}{[u_0+u]},\\
&&
K_d(\zeta;r_d)=\sum_{\nu=\pm 1}t^{*\,new}_{\ve}(u_0-u)^{l}_{l+\nu} t'^{new}_{\ve'}(u_0+u)^{l+\nu}_l K^{(c)}_<\Kmatrix{l+\nu}{l}{l}{u},\\
&&
\qquad\qquad=\frac{f\!\left(1-\frac r2-\frac{i\pi}{2\epsilon}+u\right)}{f\!\left(1-\frac r2-\frac{i\pi}{2\epsilon}-u\right)}K\!\left(u;c,l={r\over 2}-c,u_0=1-{r\over 2}-{i\pi\over 2\epsilon}\right).
\ena
But these new intertwining vectors satisfy the relations \eqref{eqn:vec-rel} as before and in the formuale for the correlation function \eqref{eqn:L'}, \eqref{eqn:phiphi} the vectors always consist such a pair that the factor $f(u)$'s cancel. Therefore we can substitute the vectors dressed with $f(u)$ for those not dressed.


It can be verified that this $K(u;c,l,u_0)$ coincides with $K_g(u_g;\xi_g,\lambda_g,\mu_g)$: the general solution  of the boundary Yang-Baxter equation obtained in \cite{IK} (We use the subscript $g$ for \cite{IK}). Relations between the parameters $\{c,l,u_0\}\leftrightarrow \{\xi_g,\lambda_g,\mu_g\}$ are highly complicated and discussed in Appendix \ref{appdx:IK}.


\subsection{Correlation functions}
In this section, we give the integral formulae for the correlation functions of the boundary XYZ model. For clarity we mainly discuss a special case of one-point functions i.e. the boundary magnetization but generalization to $N$-point case is straightforward though cumbersome.



\noindent {\bf Basic idea}\\
We denote the matrix unit operator acting on the site $r$ as $E^{r}_{\ve\ve'}$ e.g.
\[
E^{r}_{\pm\mp}=\sigma^\pm_r,\quad E^{r}_{++}-E^{r}_{--}=\sigma^z_r.
\]
Our strategy is represented as the following naive equation
\bea
&&
{}_B\bra{0}E^{1}_{\ve'\ve}\ket{0}_B\non\\
&&
=(\mbox{The partition function of Fig.\ref{fig:latt-v}})\non\\
&&
=\lim_{\xi,\xi'\to 1}{}_B\bra{0}\phi^*_{\ve'}(\xi')\phi_{\ve}(\xi)\ket{0}_B\lb{eqn:corr-v}\non\\[2mm]
&&
=\lim_{\xi,\xi'\to 1}(\mbox{Fig.}\ref{fig:ctm-v})\non\\
&&
=\lim_{\xi,\xi'\to 1}(\mbox{Fig.}\ref{fig:ctm-f})
\lb{eqn:corr-fn}\\
&&
=\lim_{\xi,\xi'\to 1}\sum_{s,s'}{}^{\ c}_B\bra{l-1,l}\phi^*(v')^l_{s'}\phi(v)^{s'}_s\Lambda'(u_0)^s_{l}\ket{l-1,l}^c_B\times t^*_{\ve'}(u_0-v)^l_{s'}t_{\ve}(u_0-v)^{s'}_{s} ,\non\\
\mbox{where}&&\non
\\
&&
\xi=x^{2v},\quad \xi'=x^{2v'}.
\ena
The first equality comes from \eqref{eqn:Hamil-ren} and the same reasoning as in the bulk case \cite{Bax}, \cite{JM}.\\
\noi The second and the third equalities are argued in Sec.3 of \cite{bdry qKZ}. They argued the equivalence of two lattices Fig.\ref{fig:latt-v} and Fig.\ref{fig:ctm-v} and made following identification
\bea
&&
\ket{0}_B\sim (\mbox{Upper half lattice of Fig.\ref{fig:latt-v}}),\\
&&
{}_B\bra{0}\sim (\mbox{Lower half lattice of Fig.\ref{fig:latt-v}}).
\ena   
The fourth equality is from the equivalence of two lattices in Fig.\ref{fig:fv-latt}. Iteration of face-vertex correspondences of the local Boltzmann weights on one lattice yields the other. This is the boundary version of \cite{LaP}.\\  
\noi The fifth equality comes from the similar augument for the second and third above.  

Thus the one-point function for the XYZ model is equivalent to the two-point function of the boundary SOS model with the insertion of the tail operator $\Lambda'(u)^m_n$. Here $m=n-2k,\:(k\in\mathbb{Z})$ because the number of vertex operator is even in \eqref{eqn:corr-fn}. Graphical argument shows that the tail operator satisfies the commutation relation
\bea
&&
\Lambda'(u_0)^{n'}_s\phi^*(u)^s_n=\sum_{s'}L'\LL{n}{s}{s'}{n'}{u_0-u} \phi^*(u)^{n'}_{s'}\Lambda'(u_0)^{s'}_n,\\
&&
L'\LL{n'}{s'}{s}{n}{v}=\sum_{\ve=\pm 1}t_{\ve}'(v)^{n'}_{s'}t_{\ve}^*(v)^{n}_{s}\label{eqn:L'}.
\ena
To bosonise $\Lambda'(u_0)^{n-2k}_n$ we have to consider two cases $k\leq 0$ and $k\geq 0$ separately \cite{LaP}. In the former case, we use the bosonisation of Sec.\ref{subsec:boson}, \ref{sec:bvs} and the tail operator is given by
\bea
&&
\Lambda'(u)^{n-2k}_n=\Lambda'(u)^{-2k}=\Lambda(-u)^{-2k}(-)^K,\qquad (k\in\mathbb{N})\\[2mm]
&&
\Lambda(u)^{-2k}=X_-^k(u){[K-2k]\over [K]}(-)^{(K-1)k},
\lb{eqn:LX}\\
&&
X_-(u)=\oint {d\z_1\over 2\pi i\z_1} \;x_-(u_1){[u-u_1-1/2+K] \over [u-u_1+1/2]}.
\ena
Note that
\[
\Phi_+(u)=\Phi_-(u)X_-(u).
\]
The commutation relation on a Fock space is
\bea
\Lambda'(u_0)^{-2k}\bPhi^*_\ve(-u)=\sum_{\ve'=\pm}L'\LL{K+2k-\ve}{K+2k}{K-\ve'}{K}{u_0-u} \bPhi^*_{\ve'}(-u)\Lambda'(u_0)^{-2k-\ve'+\ve}.
\ena
For $k<0$, \eqref{eqn:LX} is meaningless and we use the bosonisation of Sec.\ref{subsec:another-boson} and
\bea
\Lambda'(u)^{n-2k}_n=\Lambda'(u)^{2k}=\Lambda(-u)^{2k}(-)^K,\qquad (-k\in\mathbb{N}).
\ena
Normaizing $\Lambda'(u)^m_n$ is needledss for our purpose, see e.g. \eqref{eqn:M}.

\noindent {\bf $K$-matrix of general type}\\
We evaluate the correlation with two types of bosonization in Sec.\ref{sec:SOS} as
\bea
&&
{}_B\bra{0}\phi^*_{\ve'}(\xi)\phi_{\ve}(\xi)\ket{0}_B\non\\[2mm]
&&
=\sum_{ {s'=l\pm 1} \atop {s=s'\pm 1} }{}^{\ c}_B\bra{l-1,l}\phi^*(v)^l_{s'}\phi(v)^{s'}_s\Lambda'(u_0)^s_{l}\ket{l-1,l}^c_B\times t^*_{\ve'}(u_0-v)^l_{s'}t_{\ve}(u_0-v)^{s'}_{s}\non\\
&&
=g\ (-)^l\ {}^{\ c}_B\bra{l-1,l}\bPhi^*_+(-v)\bPhi_+(-v)\ket{l-1,l}^c_B\times t^*_{\ve'}(u_0-v)^{l}_{l-1}t_{\ve}(u_0-v)^{l-1}_{l}\non\\[2mm]
&&
+g\ {}^{\ c}_B\bra{l-1,l}\bPhi^*_+(-v)\bPhi_-(-v)\Lambda'^{-2}(u_0)\ket{l-1,l}^c_B\times t^*_{\ve'}(u_0-v)^{l}_{l-1}t_{\ve}(u_0-v)^{l-1}_{l-2}\non\\[2mm]
&&
+g\ (-)^l\ {}^{\ c}_B\bra{\overline{l-1,l}}\bPhi^*_+(-v)\bPhi_+(-v)\ket{\overline{l-1,l}}^c_B\times t^*_{\ve'}(u_0-v)^{l}_{l+1}t_{\ve}(u_0-v)^{l+1}_{l}\non\\[2mm]
&&
+g\ {}^{\ c}_B\bra{\overline{l-1,l}}\bPhi^*_+(-v)\bPhi_-(-v)\Lambda'^{-2}(u_0)\ket{\overline{l-1,l}}^c_B\times t^*_{\ve'}(u_0-v)^{l}_{l+1}t_{\ve}(u_0-v)^{l+1}_{l+2}.\lb{eqn:phiphi}\non\\
&&
\ena
The boundary vacuum expectation values in this formula are one-fold integrals and the essential part of the integrands is
\bea
&&
{
{}^c_B\langle k-1,k|:\Phi_-(-v-1)\Phi_-(-v)x_-(u_1):|k-1,k \rangle^c_B
\over
{}^c_B\langle k-1,k|k,k-1 \rangle^c_B
}\non\\[2mm]
&&
=\left(x\xi z_1\right)^{{l\over r}-1}\left(z_1^2\over x\xi\right)^{{\rs\over 2r}}{(x^2;x^2)_\infty^3\over (x^2)_\infty^2(x^4;x^4)_\infty^2}\non\\
&&
\times{\Theta_{x^{2r}}(x^{2(c+r)}/\xi)\Theta_{x^{2r}}(x^{2(c+k)}\xi)\Theta_{x^4}(\xi^{-2})\over \Theta_{x^{2r}}(\xi^{-2})}\non\\
&&
\times{\Theta_{x^4}(z_1^{-2})\over \Theta_{x^{2r}}(x^{-2c-1}/z_1)_\infty\Theta_{x^{2r}}(x^{2(c+k)-1}/z_1)_\infty}\non\\
&&
\times{\Theta_{x^{2r}}(xz_1/\xi)\ (x^3\xi z_1)_\infty(x\xi^{-1}z_1^{-1})_\infty\over\Theta_{x^2}(x\xi/z_1)\ (x^3\xi z_1;x^2)_\infty(x\xi^{-1}z_1^{-1};x^2)_\infty}.
\ena
Then the boundary magnetization with a spectral parameter is given as
\bea
&&
\frac{{}_B\bra{0}\sigma_1^z\ket{0}_B}{{}_B\br{0|0}_B}
=\lim_{\xi\to 1}\M^{(0)}(\xi;c,l,u_0)
\ena
We used the coherent states to obtain this formula, see \cite{MW} for detail.
\bea
&&
\M^{(0)}(\xi;c,l,u_0)=\sum_{\ve=\pm}\frac{\ve\, {}_B\bra{0}\phi^*_{\ve}(\xi)\phi_{\ve}(\xi)\ket{0}_B}{{}_B\br{0|0}_B}\\[2mm]
&&
\qquad =\sum_{ {s'=l\pm 1} \atop {s=s'\pm 1} }{{}^{\ c}_B\bra{l-1,l}\phi^*(v)^l_{s'}\phi(v)^{s'}_s\Lambda'(u_0)^s_{l}\ket{l-1,l}^c_B \over {}^{\ c}_B\br{l-1,l|\Lambda'(u_0)^l_l|l-1,l}^c_B}\sum_{\ve=\pm} \ve\, t^*_{\ve}(u_0-v)^l_{s'}t_{\ve}(u_0-v)^{s'}_{s}\non\\
&&\lb{eqn:M}\\
&&
\qquad =F(\xi;c,l,u_0)+F(\xi;-r-c,2r-l,u_0),
\ena
\bea
&&
F(\xi;c,l,u_0)\non\\[2mm]
&&
=g\, 
{
 {}^{\ c}_B\bra{l-1,l}\bPhi^*_+(-v)\bPhi_+(-v)\ket{l-1,l}^c_B 
\over
 {}^{\ c}_B\br{l-1,l|l-1,l}^c_B
}
\times \sum_{\ve=\pm} \ve\, t^*_{\ve}(u_0-v)^{l}_{l-1}t_{\ve}(u_0-v)^{l-1}_{l}
\lb{eqn:two-term}\\[2mm]
&&
\quad +g\ 
{
 {}^{\ c}_B\bra{l-1,l}\bPhi^*_+(-v)\bPhi_-(-v)\Lambda'^{-2}(u_0)\ket{l-1,l}^c_B 
\over 
 (-)^l\: {}^{\ c}_B\br{l-1,l|l-1,l}^c_B
}
\times \sum_{\ve=\pm} \ve\, t^*_{\ve}(u_0-v)^{l}_{l-1}t_{\ve}(u_0-v)^{l-1}_{l-2}\non\\[2mm]
&&
=-g\,{\dbr{\Phi_-(-v-1)\Phi_-(-v)} \over [l]}\oint_\C{d\z_1\over 2\pi i\z_1}{h_4(u_1+u_0-1/2)h_4(l-v-u_1-1/2) \over [-v-u_1+1/2][-u_0-u_1+1/2]}\non\\[2mm]
&&
\quad\times {{}^{\ c}_B\langle l-1,l|:\Phi_-(-v-1)\Phi_-(-v)x_-(u_1):|l,l-1 \rangle^c_B \over {}^{\ c}_B\langle l-1,l|l,l-1 \rangle^c_B}\non\\[2mm]
&&
\quad\times\dbr{\Phi_-(-v-1)x_-(u_1)}\dbr{\Phi_-(-v)x_-(u_1)},\\[2mm]
&&
=\oint_\C{d\z_1\over 2\pi i\z_1}F(\xi,\z_1;c,l,u_0)
\ena
And
\bea
&&
F(\xi,\z_1;c,l,u_0)\non\\
&&
=-{1\over [l]}{\Theta_{x^{2r}}(x^{2(c+r)}/\xi)\Theta_{x^{2r}}(x^{2(c+l)}\xi)\Theta_{x^{4}}(\xi^{-2}) \over \Theta_{x^{2r}}(\xi^{-2})}\non\\[2mm]
&&
\quad\times{h_4(u_1+u_0-1/2)h_4(l-v-u_1-1/2) \over [-v-u_1+1/2][-u_0-u_1+1/2]}(x\xi \z_1)^{l/r}x^{-1}(\xi \z_1)^{-1/r}\non\\[2mm]
&&
\quad\times {\Theta_{x^{2r}}(x\z_1/\xi)\Theta_{x^{2r}}(x\z_1^{-1}\xi^{-1}) \over \Theta_{x^{2}}(x\xi/\z_1)\Theta_{x^{2}}(x\xi \z_1)}{\Theta_{x^{4}}(\z_1^{-2}) \over \Theta_{x^{2r}}(x^{-2c-1}/\z_1)\Theta_{x^{2r}}(x^{2(c+l)-1}/\z_1)}\non\\[2mm]
&&
\quad\times {(x^2;x^2)_\infty^4 \over (x^4;x^4)_\infty^2}.
\ena
Where $\dbr{\ldots}$ are the factors for normal ordering and given in Appendix \ref{appdx:OPE}
 and
\bea
&&
h_1(u)=
C\vartheta_1\!\left({u\over r};{i\pi\over \epsilon r}\right)
=[u]=\mbox{fr}_1(u)\Theta_{x^{2r}}(x^{2u}),
\quad\mbox{fr}_1(u)=x^{{u^2\over r}-u}\\[2mm]
&&
h_4(u)=
C\vartheta_4\!\left({u\over r};{i\pi\over \epsilon r}\right)
=\mbox{fr}_4(u)\left[u-{i\pi\over 2\epsilon}\right],
\quad\mbox{fr}_4(u)=e^{-{\pi^2\over 4r \epsilon}-{i\pi u\over r}}.
\ena
The cotour $\C$ is such that 

$\z_1=x^{2n+1}\xi,x^{2n+1}\xi^{-1},x^{2rn-2c-1},x^{2rn+1}\z_0^{-1},x^{2rn+2(c+l)-1}\;(n\geq0)$

\noindent are inside and 

$\z_1=x^{2n-1}\xi,x^{2n+1}\xi^{-1},x^{2rn-2c-1},x^{2rn+1}\z_0^{-1},x^{2rn+2(c+l)-1}\;(n<0)$
 
\noindent are outside of it. Note that $\z_1=x^{-1}\xi$ is not a pole of $F(\xi,\z_1;c,l,u_0)$.

We used following identities to make two terms in \eqref{eqn:two-term} together
\be
&&
\vartheta_4(2x)\vartheta_4(2y)\vartheta_1(2z)\vartheta_1(2t)\\
&&
=\vartheta_4(x+y+z+t)\vartheta_4(x+y-z-t)\vartheta_1(x-y-z+t)\vartheta_1(x-y+z-t)\\
&&
\quad-\vartheta_4(x+y+z-t)\vartheta_4(x+y-z+t)\vartheta_1(x-y+z+t)\vartheta_1(x-y-z-t)
\en
and
\be
\sum_{\ve=\pm}\ve\ t^*_{\ve}(u)^{n_2}_{n_1}t_{\ve}(u)^{n_4}_{n_3}={(-)^{n_1+n_3}(n_1-n_2)\over [n_1]}K\LL{n_4}{n_3}{n_1}{n_2}{u},
\en
where
\bea
&&
K\LL{n'}{n'\pm 1}{n}{n\pm 1}{u}=
{
h_4(u\pm {n-n'\over 2}) h_4({n+n'\over 2})
\over
h_1(u)
},\\
&&
K\LL{n'}{n'\mp 1}{n}{n\pm 1}{u}={h_4(u\pm {n+n'\over 2})h_4({n-n'\over 2})\over h_1(u)}.\\[2mm]\ena

\noindent {\bf Diagonal $K$-matrix} \\
From \eqref{eqn:cond-diag}, for the case of diagonal $K$-matirx discussed in \cite{bdry qKZ} we have 
\bea
&&
\M^{(0)}(\xi;c)=\M^{(0)}\left(\xi;c,l={r\over 2}-c,u_0=1-{r\over 2}-{i\pi\over 2\epsilon}\right)\\
&&
\qquad\qquad\; =F(\xi;c)+F(\xi;-r-c)\\
\mbox{where} && \non\\
&&
F(\xi;c)=F\left(\xi;c,l={r\over 2}-c,u_0=1-{r\over 2}-{i\pi\over 2\epsilon}\right)\\
&&
\qquad\quad=\oint_{\C'}{d\z_1\over 2\pi i\z_1}F(\xi,\z_1;c)\\
&&
F(\xi,\z_1;c)=F\left(\xi,\z_1;c,l={r\over 2}-c,u_0=1-{r\over 2}-{i\pi\over 2\epsilon}\right)\\
&&
\qquad\qquad={1\over [{r\over 2}-c]}{\Theta_{x^{2r}}(x^{2(c+r)}/\xi)\Theta_{x^{2r}}(x^{r}\xi)\Theta_{x^{4}}(\xi^{-2}) \over \Theta_{x^{2r}}(\xi^{-2})}\non\\[2mm]
&&
\qquad\qquad\times{h_4\!\left(\frac{r-1}{2}-c-v-u_1\right) \over\left[-u_1-\frac{1-r}{2}+\frac{i\pi}{2\epsilon}\right]}
{\Theta_{x^{2r}}(x\z_1/\xi)\over \Theta_{x^{2}}(x\xi/\z_1)\Theta_{x^{2}}(x\xi \z_1)}{\Theta_{x^{4}}(\z_1^{-2}) \over \Theta_{x^{2r}}(x^{-2c-1}/\z_1)}\non\\
&&
\qquad\qquad \times x^{-\frac 12-\frac cr-r}\xi^{\frac 12-\frac{1+c}{r}}\z_1^{\frac 32-\frac{1+c}{r}}\frac{\mbox{fr}_4\!\left(u_1+\frac{1-r}{2}-\frac{i\pi}{2\epsilon}\right)\mbox{fr}_1\!\left(u_1+\frac{1-r}{2}-\frac{i\pi}{\epsilon}\right)}{\mbox{fr}_1\!\left(-v-u_1+\frac 12\right)}\non\\
&&
\qquad\qquad\times {(x^2;x^2)_\infty^4 \over (x^4;x^4)_\infty^2}
\ena
the contour $\C'$ is such that 

$\z_1=x^{2n+1}\xi,x^{2n+1}\xi^{-1},x^{2rn-2c-1},-x^{2rn+r-1}\:(n\geq0)$

\noindent are inside and 

$\z_1=x^{2n-1}\xi,x^{2n+1}\xi^{-1},x^{2rn-2c-1},-x^{2rn+r-1}\:(n<0)$

\noindent are outside of it. As before $\z_1=x^{-1}\xi$ is not a pole of $F(\xi,\z_1;c,l,u_0)$.
We can reproduce the result of \cite{bdry qKZ} as a special case of these formulae
\bea
&&
\M^{(0)}\left(\xi;c={i\pi\over 2\epsilon}\right)
=\mbox{Res}\left(\z_1=-x^{-1};F\left(\xi,\z_1;c={i\pi\over 2\epsilon}\right){d\z_1\over 2\pi i\z_1}\right)\\[2mm]
&&
\qquad\qquad={(x^2\xi^{-1};x^2)_\infty(x^2\xi;x^2)_\infty \over (-x^2\xi^{-1};x^2)_\infty(-x^2\xi;x^2)_\infty}{(-x^{2r}\xi^{-1})_\infty(-x^{2r}\xi)_\infty \over (x^{2r}\xi^{-1})_\infty(x^{2r}\xi)_\infty}{(x^2;x^2)_\infty^2(-x^{2r})_\infty^2 \over (-x^2;x^2)_\infty^2(x^{2r})_\infty^2},\non\\
&&\\
&&
(z)_\infty=(z;x^{2r})_\infty,
\ena
where we used
\bea
F\left(\xi,\z_1;c={i\pi\over 2\epsilon}\right)=-F\left(\xi,\z_1;c=-r-{i\pi\over 2\epsilon}\right)
\ena

The formula for the difference of the boundary magnetizations in \cite{bdry qKZ} can be also reproduced. We have
\bea
&&
\M^{(0)}(\xi;c)=\M^{(0)}(\xi^{-1};c),\\
&&
\M^{(0)}(\xi;c)=-\M^{(1)}(\xi;-c),
\ena
which can be seen easily by physical argument and discussed rigorously in \cite{bdry XXZ}. 
hence
\bea
\M^{(0)}(\xi;c)-\M^{(1)}(\xi;c)=\M^{(0)}(\xi;c)+\M^{(0)}(\xi^{-1};-c)
\ena
and R.H.S can be written in a simple form 
\bea
&&
\M^{(0)}(\xi;c)+\M^{(0)}(\xi^{-1};-c)=\mbox{Res}\left(\z_1=x^{-2c-1};F\left(\xi,\z_1;c\right){d\z_1\over 2\pi i\z_1}\right)\non\\[2mm]
&&
\hspace{5cm}-\mbox{Res}\left(\z_1=x^{-1}\xi^{-1};F\left(\xi,\z_1;c\right){d\z_1\over 2\pi i\z_1}\right)\non\\[2mm]
&&
\hspace{5cm}-\mbox{Res}\left(\z_1=x^{2c-1};F\left(\xi,\z_1;-r-c\right){d\z_1\over 2\pi i\z_1}\right)\non\\[2mm]
&&
\hspace{5cm}-\mbox{Res}\left(\z_1=x^{-1}\xi^{-1};F\left(\xi,\z_1;-r-c\right){d\z_1\over 2\pi i\z_1}\right)
\ena
where we used
\bea
&&
F(\xi^{-1},\z_1;-c){d\z_1\over 2\pi i\z_1}=F(\xi,w_1;c){dw_1\over 2\pi iw_1}\\
&&
F(\xi^{-1},\z_1;-r+c){d\z_1\over 2\pi i\z_1}=F(\xi,w_1;-r-c){dw_1\over 2\pi iw_1}\\&&
\qquad w_1=\frac{1}{x^2\z_1}.
\ena
Therefore  
\bea
&&
\M^{(1)}(1;c)-\M^{(0)}(1;c)=-2\mbox{Res}\left(\z_1=x^{-2c-1};F\left(1,\z_1;c\right){d\z_1\over 2\pi i\z_1}\right)\\[2mm]
&&
\qquad =2\frac{(x^r)_\infty^2(-x^r)_\infty^2(x^{2r+2c})_\infty^2(x^{-2c})_\infty^2}{(x^{r+2c})_\infty(-x^{r+2c})_\infty(x^{r-2c})_\infty(-x^{r-2c})_\infty}\non\\
&&
\qquad\qquad
\times\frac{(x^{4c+2};x^4)_\infty(x^{-4c+2};x^4)_\infty}{(x^{2+2c};x^2)_\infty^2(x^{-2c};x^2)_\infty^2}\frac{(x^4;x^4)_\infty^2(x^2;x^2)_\infty^2}{(x^{2r})_\infty^4}.
\ena
Note that we are taking $\ket{0}_B$ as the ground state.
We can reproduce the result of \cite{bdry XXZ} by taking limit $x^{2r}\to 0$
\bea
\lim_{x^{2r}\to 0}\left(\M^{(1)}(1;c)-\M^{(0)}(1;c)\right)=2\frac{(x^{4c+2};x^4)_\infty(x^{-4c+2};x^4)_\infty}{(x^{2+2c};x^2)_\infty^2(x^{2-2c};x^2)_\infty^2}(x^4;x^4)_\infty^2(x^2;x^2)_\infty^2,
\ena
where the parameter $r$ of \cite{bdry XXZ} is identified with $x^{-2c}$.

\noi {\bf $N$-point correlation function}\\
The $N$-point correlation function can be obtained in the same manner as discussed in \cite{JM}, \cite{bdry XXZ}. But the formula is highly complicated. Here we write down only the essential part.
\bea
&&
{}_B\bra{0}\phi^*_{\ve_1'}(\xi_1')\ldots\phi^*_{\ve_N'}(\xi_N')\phi_{\ve_N}(\xi_N)\ldots\phi_{\ve_1}(\xi_1)\ket{0}_B\non\\[2mm]
&&
=\sum_{s_1'\ldots s_N'\atop s_1\ldots s_N}{}^{\ c}_B\bra{l-1,l}
\phi^*(v_1')^l_{s_1'}\phi^*(v_2')^{s_1'}_{s_2'}\ldots\phi^*(v_N')^{s_{N-1}'}_{s_N'}\non\\
&&
\hspace{6cm}\times
\phi(v_N)^{s_N'}_{s_N}\phi(v_{N-1})^{s_N}_{s_{N-1}}\ldots\phi(v_1)^{s_2}_{s_1}\Lambda'(u_0)^{s_1}_{l}\ket{l-1,l}^c_B\non\\[2mm]
&&
\hspace{3cm}\times
t^*_{\ve_1'}(u_0-v_1')^{l}_{s_1'}t^*_{\ve_2'}(u_0-v_2')^{s_1'}_{s_2'}\ldots t^*_{\ve_N'}(u_0-v_N')^{s_{N-1}'}_{s_N'}\non\\
&&
\hspace{3cm}\times 
t_{\ve_N}(u_0-v_N)^{s_{N}'}_{s_N}t_{\ve_{N-1}}(u_0-v_{N-1})^{s_{N}}_{s_{N-1}}\ldots t_{\ve_1}(u_0-v_1)^{s_2}_{s_1}\\[2mm]
&&
=g\!\!\sum_{\nu_i,\nu_i'\atop \nu_i-\nu_i'\leq 0}{}^{\ c}_B\bra{l-1,l}
\bPhi^*_{\nu_1'}(-v_1')\ldots\bPhi^*_{\nu_N'}(-v_N')
\bPhi_{\nu_N}(-v_N)\ldots\bPhi_{\nu_1}(-v_1)
\Lambda'^{\sum_{i=1}^N\nu_i-\nu_i'}(u_0)
\ket{l-1,l}^c_B\non\\[2mm]
&&
\hspace{3cm}\times
\prod_i
t^*_{\ve_i'}(u_0-v_i')^{ l-\sum_{n=1}^{i-1}\nu_n' }_{ l-\sum_{n=1}^{i}\nu_n' }
t_{\ve_i}(u_0-v_i)^{ l-\sum_{m=1}^{N}\nu_m'+\sum_{n=N}^{j+1}\nu_n }_{ l-\sum_{m=1}^{N}\nu_m'+\sum_{n=N}^{j}\nu_n}\non\\[2mm]
&&
+g\!\!\sum_{\nu_i,\nu_i'\atop \nu_i-\nu_i'< 0}{}^{\ c}_B\bra{\overline{l-1,l}}
\bPhi^*_{\nu_1'}(-v_1')\ldots\bPhi^*_{\nu_N'}(-v_N')
\bPhi_{\nu_N}(-v_N)\ldots\bPhi_{\nu_1}(-v_1)
\Lambda'^{\sum_{i=1}^N\nu_i-\nu_i'}(u_0)
\ket{\overline{l-1,l}}^c_B.\non\\
&&
\hspace{3cm}\times
\prod_i
t^*_{\ve_i'}(u_0-v_i')^{ l+\sum_{n=1}^{i-1}\nu_n' }_{ l+\sum_{n=1}^{i}\nu_n' }
t_{\ve_i}(u_0-v_i)^{ l+\sum_{m=1}^{N}\nu_m'-\sum_{n=N}^{j+1}\nu_n }_{ l+\sum_{m=1}^{N}\nu_m'-\sum_{n=N}^{j}\nu_n}.
\ena
Note that each term in R.H.S is $N$-fold integral.


\setcounter{section}{4}
\setcounter{equation}{0}
\section{Discussion}
\lb{sec:discuss}
In \cite{LaP}, \lq\lq m and $u_0$''-independence is carefully discussed. We explain this point with Fig.\ref{fig:fv-latt}. For the bulk problem, the correlation functions of the eight-vertex model should not depend on the parameters of the sorrounding intertwining vectors ($l$ and $u_0$) in the thermodynamic limit. (We use $l$ for $m$ of \cite{LaP}.) This is nontrivial because the essential part of the correlation function has the form
\bea
\tr_{\F_{l,n_2}}(\Phi(u-1)^{n_2}_{n_1}\ldots\Lambda(u_0)x^{4H_n}),
\ena
where $H_n$ is the corner transfer hamiltonian of the ABF model (see (5.14) of \cite{LaP}).
In our case, even in the thermodynamic limit $l$ and $u_0$ remain as the parameters of the $K$-matrix. Therfore the correlation functions should depend on them as they do. 

It is argued in \cite{FHS} that the general solution of the reflection equation of the ABF model can be constructed from that of the eight-vertex model through the face-vertex correspondence. Combining this with the argument of Sec.\ref{subsec:fv-K},  for the ABF model we can construct the general solution from the diagonal $K$-matrix. But this method is not applicable for the eight-vertex model since in our construction of \eqref{eqn:fv-K} we can not use a non-diagonal $K$-matrix of the ABF model. 

Finally we want to mention the related work of \cite{FHSY}. In this paper the Bethe ansatz equation is obtained for the eight-vertex model with two-sided boundaries. The $K$-matrix considered is a general solution of the boundary Yang-Baxter equation.

\vspace{1cm}
\noindent
{\bf {\Large Acknowledgments}}\\
The author thanks Hitoshi Konno, Satoru Odake, Yaroslav Pugai, Jun'ichi Shiraishi and Junji Suzuki for discussions and interest. 
He is especially indebted to Michio Jimbo and  Atsuo Kuniba for their advice and encouragement.
%
%

\appendix


\setcounter{equation}{0}
\section{Notations and definitions of functions}
\lb{appdx:fn}

\noi{\bf Elliptic functions}\\
We follow Chap.15 of \cite{Bax} with some modifications. $\vartheta_i(u;\tau)$'s are given as
\bea
&&
\vartheta_1(u;\tau)=H(2Iu),\quad \vartheta_2(u;\tau)=H_1(2Iu),\\
&&
\vartheta_3(u;\tau)=\Theta(2Iu),\quad \vartheta_4(u;\tau)=\Theta_1(2Iu),\\
&&
e^{i\pi\tau}=q.
\ena
where R.H.S.'s are of \cite{Bax}.
$k,k'$ are elliptic moduluses and for half-period magnitudes we use $K,K'$ instead of $I,I'$.
sn$(u,k)$, cn$(u,k)$ and dn$(u,k)$ are the same as in \cite{Bax}.
We also use the followings
\bea
{\rm snh}(u,k)=-i{\rm sn}(iu,k),\quad {\rm cnh}(u,k)={\rm cn}(iu,k),\quad {\rm dnh}(u,k)={\rm dn}(iu,k).
\ena

For convenience, we gather some notations for elliptic functions
\bea
&&
h_1(u)=
C\vartheta_1\!\left({u\over r};{i\pi\over \epsilon r}\right)
=[u]=\mbox{fr}_1(u)\Theta_{x^{2r}}(x^{2u}),
\quad\mbox{fr}_1(u)=x^{{u^2\over r}-u},\\[2mm]
&&
h_4(u)=
C\vartheta_4\!\left({u\over r};{i\pi\over \epsilon r}\right)
=\mbox{fr}_4(u)\left[u-{i\pi\over 2\epsilon}\right],
\quad\mbox{fr}_4(u)=e^{-{\pi^2\over 4r \epsilon}-{i\pi u\over r}},\\[2mm]
&&
\Theta_p(z)=(z;p)_\infty(pz^{-1};p)_\infty(p;p)_\infty,\\[2mm]
&&
C=\sqrt{{\pi\over\epsilon r}}e^{\epsilon r/4}.
\ena

\noi{\bf Other functions}\\
\bea
&&
\{z\}=(z;x^4,x^{2r})_\infty,\\
&&
(z)_\infty=(z;x^{2r})_\infty,\\
&&
(z;p_1,\ldots,p_N)_\infty
=\prod_{n_1,\ldots,n_N=0}^\infty(1-zp_1^{n_1}\ldots p_N^{n_N}).
\ena


\setcounter{equation}{0}
\section{Formulae for normal ordering}
\lb{appdx:OPE}
We list the formulae for normal ordering. For operators $A,B$, we write down $\dbr{AB}$ such that $AB=\dbr{AB}\times :AB:$.
\bea
&&
\dbr{ x_-(u_1) x_-(u_2) }=
z_1^{-{2\over r}+2}(1-{z_2\over z_1})
{(x^{ 2}z_2/z_1)_\infty \over (x^{2r-2}z_2/z_1)_\infty},\\
&&
\dbr{ \Phi_-(u_1)x_-(u_2) }
=z_1^{{1\over r}-1 } 
{(x^{ 2r-1 }z_2/z_1)_\infty \over (x z_2/z_1)_\infty},\\
&&
\dbr{ x_-(u_2)\Phi_-(u_1) }
=z_2^{{1\over r}-1 } 
{(x^{ 2r-1 }z_1/z_2)_\infty \over (x z_1/z_2)_\infty},\\
&&
\dbr{ \Phi_-(u_1)\Phi_-(u_2) }=
z_1^{\rs\over 2r} 
{ \{x^2 z_2/z_1\} \{x^{2r+2} z_2/z_1\} \over
  \{x^4 z_2/z_1\} \{x^{2r } z_2/z_1\} }.
\ena
As meromorphic functions, following commutation relations hold
\bea
&&
x_-(u_1) x_-(u_2) = {[u_1-u_2-1]  \over [u_1-u_2+1] }
   x_-(u_2) x_-(u_1),\\
&&
 \Phi_-(u_1)x_-(u_2)=
-{[u_1-u_2+1/2]  \over [u_1-u_2-1/2] }x_-(u_2) \Phi_-(u_1),\\
&&
 \Phi_-(u_1)  \Phi_-(u_2) =
R_0(u_1-u_2) \Phi_-(u_2)  \Phi_-(u_1) .
\ena


\setcounter{equation}{0}
\section{Correspondence of the $K$-matrix with \cite{IK}}
\lb{appdx:IK}
As claimed in Sec.\ref{subsec:fv-K}, we detail how $K(u;c,l,u_0)$ coincides with the $K$-matrix given in \cite{IK}. The subscript $g$ is for those of \cite{IK} which is a general solution of the boundary Yang-Baxter eauation. The relations between parameters are given as
\bea
&&
i\frac{K'_g}{K_g}=-\frac{2}{\tau_0},\\
&&
\frac{u_g}{2K_g}=\frac{u}{r\tau_0},\\
&&
\frac{\eta_g}{2K}-1=-\frac{1}{r\tau_0},\\
&&
k=k_g,\quad\mbox{(elliptic modulus)}\\
\mbox{where}&&\non\\
&&
\tau_0=\frac{i\pi}{\epsilon r}.
\ena
In [IK] $K_g(u_g;\xi_g,\lambda_g,\mu_g)$ is given as
\bea
&&
K_g(u_g;\xi_g,\lambda_g,\mu_g)^\pm_\pm=\frac{\sn(\xi_g\pm u_g;k_g)}{\sn(\xi_g;k_g)},\\
&&
K_g(u_g;\xi_g,\lambda_g,\mu_g)^\pm_\mp=\mu_g\frac{\sn(2u_g;k_g)}{\sn(\xi_g;k_g)}\frac{{\dstyle \lambda_g(1-k_g\sn^2(u_g;k_g))\mp 1\mp k_g\sn^2(u_g;k_g)}}{{\dstyle 1-k_g^2\sn^2(\xi_g;k_g)\sn^2(u_g;k_g)}} .\\
&&
\ena
We rewrite it as
\bea
&&
K_g(u_g;\xi_g,\lambda_g,\mu_g)\non\\
&&
=\left\{
\sn(\xi_g;k_g)(1-k_g^2\sn^2(u_g;k_g)\sn^2(\xi_g;k_g))
\vartheta_0^2\!\left(\frac{u_g}{2K_g};-\frac{2}{\tau_0}\right)
\vartheta_0^2\!\left(\frac{\xi_g}{2K_g};-\frac{2}{\tau_0}\right)
\vartheta_0\!\left(\frac{u_g}{K_g};-\frac{2}{\tau_0}\right)
\right\}^{-1}\non\\[2mm]
&&
\qquad\times{\bar K}_g(u_g;\xi_g,\lambda_g,\mu_g) ,\\[2mm]
&&
{\bar K}_g(u_g;\xi_g,\lambda_g,\mu_g)^\pm_\pm=
c^{\enskip\,\pm}_{g,\pm}(\xi)\,
\vartheta_0\!\left(\frac{u\mp\xi}{r\tau_0};-\frac{2}{\tau_0}\right)
\vartheta_1\!\left(\frac{u\pm\xi}{r\tau_0};-\frac{2}{\tau_0}\right)
\vartheta_0\!\left(\frac{2u}{r\tau_0};-\frac{2}{\tau_0}\right),\\[2mm]
&&
{\bar K}_g(u_g;\xi_g,\lambda_g,\mu_g)^+_-=
\mu_g c^{\enskip\, +}_{g,-}(\xi,a)\,
\vartheta_1\!\left(\frac{2u}{r\tau_0};-\frac{2}{\tau_0}\right)
\vartheta_1\!\left(\frac{u-a}{r\tau_0};-\frac{2}{\tau_0}\right)
\vartheta_1\!\left(\frac{u+a}{r\tau_0};-\frac{2}{\tau_0}\right),\\[2mm]
&&
{\bar K}_g(u_g;\xi_g,\lambda_g,\mu_g)^-_+=
\mu c^{\enskip\, -}_{g,+}(\xi,a)\,
\vartheta_1\!\left(\frac{2u}{r\tau_0};-\frac{2}{\tau_0}\right)
\vartheta_0\!\left(\frac{u-a}{r\tau_0};-\frac{2}{\tau_0}\right)
\vartheta_0\!\left(\frac{u+a}{r\tau_0};-\frac{2}{\tau_0}\right),
\ena
where
\bea
&&
\frac{\xi}{r\tau_0}=\frac{\xi_g}{2K_g},\\[2mm]
&&
\lambda_g=
\frac{
\vartheta_0^2\!\left(\frac{a}{r\tau_0};-\frac{2}{\tau_0}\right)
+\vartheta_1^2\!\left(\frac{a}{r\tau_0};-\frac{2}{\tau_0}\right)
}{
\vartheta_0^2\!\left(\frac{a}{r\tau_0};-\frac{2}{\tau_0}\right)
-\vartheta_1^2\!\left(\frac{a}{r\tau_0};-\frac{2}{\tau_0}\right)
},
\ena
and
\bea
&&
c^{\enskip\, +}_{g,+}(\xi)=
2\frac{k'}{k}
\frac{
\prod_{i=1}^4 \vartheta_i\!\left(\frac{\xi}{r\tau_0};-\frac{2}{\tau_0}\right)
}{
\vartheta_0\!\left(0;-\frac{2}{\tau_0}\right)
\vartheta_1\!\left(\frac{2\xi}{r\tau_0};-\frac{2}{\tau_0}\right)
},\\[2mm]
&&
c^{\enskip\, +}_{g,-}(\xi,a)=
-2k^{-1/2}
\frac{
\vartheta_0^2\!\left(\frac{\xi}{r\tau_0};-\frac{2}{\tau_0}\right)
\vartheta_0^2\!\left(0;-\frac{2}{\tau_0}\right)
}{
\vartheta_0^2\!\left(\frac{a}{r\tau_0};-\frac{2}{\tau_0}\right)
-\vartheta_1^2\!\left(\frac{a}{r\tau_0};-\frac{2}{\tau_0}\right)
},\\[2mm]
&&
c^{\enskip\, -}_{g,-}(\xi)=-c^{\enskip\, +}_{g,+}(\xi),\quad 
c^{\enskip\, -}_{g,+}(\xi,a)=-c^{\enskip\, +}_{g,-}(\xi,a).
\ena

On the other hand ${\bar K}(u;c,l,u_0)$ of \eqref{eqn:K-bar} can be deformed into the same form
\bea
&&
{\bar K}(u;c,l,u_0)^\pm_\pm=
c^{\pm}_{\pm}(c,l,u_0)
\vartheta_0\!\left(\frac{u\mp\xi'}{r\tau_0};-\frac{2}{\tau_0}\right)
\vartheta_1\!\left(\frac{u\pm\xi'}{r\tau_0};-\frac{2}{\tau_0}\right)
\vartheta_0\!\left(\frac{2u}{r\tau_0};-\frac{2}{\tau_0}\right),\\[2mm]
&&
{\bar K}(u;c,l,u_0)^+_-=
c^{+}_{-}(c,l,u_0)
\vartheta_1\!\left(\frac{2u}{r\tau_0};-\frac{2}{\tau_0}\right)
\vartheta_1\!\left(\frac{u-a'}{r\tau_0};-\frac{2}{\tau_0}\right)
\vartheta_1\!\left(\frac{u+a'}{r\tau_0};-\frac{2}{\tau_0}\right),\\[2mm]
&&
{\bar K}(u;c,l,u_0)^-_+=
c^{-}_{+}(c,l,u_0)
\vartheta_1\!\left(\frac{2u}{r\tau_0};-\frac{2}{\tau_0}\right)
\vartheta_0\!\left(\frac{u-a'}{r\tau_0};-\frac{2}{\tau_0}\right)
\vartheta_0\!\left(\frac{u+a'}{r\tau_0};-\frac{2}{\tau_0}\right),
\ena
where
\bea
&&
\xi'=\xi'(c,l.u_0)=i\frac{\tau_0r}{2\pi}\ln R_1(l,c,u_0)\\
&&
a'=a'(l,c,u_0)=\frac{\tau_0r}{i\pi}\ln R_2(l,c,u_0),\\[2mm]
&&
R_1(l,c,u_0)=\left(\frac{C^4L^4-C^4L^2+C^2D^2L^4-C^2D^2+L^2-L}{C^4D^2L^4-C^4D^2L^2+C^2L^4-C^2+D^2L^2-D^2}\right)^{\frac 12}\\[2mm]
&&
R_2(l,c,u_0)=\frac{1}{\sqrt{2}}
\left\{
 C^2+C^{-2}+D^2+D^{-2}+C^2L^2+C^{-2}L^{-2}-C^{-2}D^{-2}L^{-2}
 \right.\\[2mm]
&&
 \left.\quad\times
 \left(
  -4C^4D^4L^4+(-C^4D^2L^4-C^4D^2L^2-C^2D^4L^2-C^2L^2-D^2L^2-D^2)^2
 \right)^{\frac 12}
\right\}^{\frac 12}\non\\
&&
C=e^{\frac{i\pi}{\tau_0r}c},\quad L=e^{\frac{i\pi}{\tau_0r}l},\quad D=e^{\frac{i\pi}{\tau_0r}\delta},
\ena
the above $C$ has nothing to do with $C$ of \eqref{eqn:C}.
And
\bea
&&
c^{+}_{+}(c,l,u_0)=\frac{2}{\tau_0}
\frac{{\dstyle
e^{-\frac{i\pi}{\tau_0 r^2}(3\xi'^2+\delta^2+2l^2+2c^2+2lc)}
}}{
\vartheta_0\!\left(0;-\frac{2}{\tau_0}\right)
\vartheta_1\!\left(\frac{2\xi'}{r\tau_0};-\frac{2}{\tau_0}\right)
\vartheta_0\!\left(\frac{2\xi'}{r\tau_0};-\frac{2}{\tau_0}\right)}\non\\[2mm]
&&
\qquad\times{\textstyle\left\{
-\vartheta_2\!\left(\frac{\xi'+\delta+l}{r\tau_0};-\frac{2}{\tau_0}\right)
\vartheta_3\!\left(\frac{\xi'-\delta+l}{r\tau_0};-\frac{2}{\tau_0}\right)
\vartheta_1\!\left(\frac{\xi'+c}{r\tau_0};-\frac{1}{\tau_0}\right)
\vartheta_1\!\left(\frac{l+c-\xi'}{r\tau_0};-\frac{1}{\tau_0}\right)
\right.}\non\\[2mm]
&&
\qquad\quad{\textstyle\left.
+\vartheta_2\!\left(\frac{\xi'+\delta-l}{r\tau_0};-\frac{2}{\tau_0}\right)
\vartheta_3\!\left(\frac{\xi'-\delta-l}{r\tau_0};-\frac{2}{\tau_0}\right)
\vartheta_1\!\left(\frac{c-\xi'}{r\tau_0};-\frac{1}{\tau_0}\right)
\vartheta_1\!\left(\frac{l+c+\xi'}{r\tau_0};-\frac{1}{\tau_0}\right)
\right\}},\\[2mm]
&&
c^{+}_{-}(c,l,u_0)=\frac{2}{\tau_0}
\frac{
(-)^l{\dstyle e^{-\frac{i\pi}{\tau_0 r^2}(\frac{3}{4}r^2+\delta^2+2l^2+2c^2+2lc)}}
}{
\vartheta_1\!\left(-\frac{1}{\tau_0};-\frac{2}{\tau_0}\right)
\vartheta_1\!\left(\frac{-r/2-a'}{r\tau_0};-\frac{2}{\tau_0}\right)
\vartheta_1\!\left(\frac{-r/2+a'}{r\tau_0};-\frac{2}{\tau_0}\right)
}\non\\[2mm]
&&
\qquad\times{\textstyle
\left\{
-\vartheta_2\!\left(\frac{-r/2+\delta+l}{r\tau_0};-\frac{2}{\tau_0}\right)
\vartheta_2\!\left(\frac{-r/2-\delta+l}{r\tau_0};-\frac{2}{\tau_0}\right)
\vartheta_1\!\left(\frac{-r/2+c}{r\tau_0};-\frac{1}{\tau_0}\right)
\vartheta_1\!\left(\frac{l+c+r/2}{r\tau_0};-\frac{1}{\tau_0}\right)
\right.}\non\\[2mm]
&&
\qquad{\textstyle\left.
+\vartheta_2\!\left(\frac{-r/2+\delta-l}{r\tau_0};-\frac{2}{\tau_0}\right)
\vartheta_2\!\left(\frac{-r/2-\delta-l}{r\tau_0};-\frac{2}{\tau_0}\right)
\vartheta_1\!\left(\frac{c+r/2}{r\tau_0};-\frac{1}{\tau_0}\right)
\vartheta_1\!\left(\frac{l+c-r/2}{r\tau_0};-\frac{1}{\tau_0}\right)\right\}},
,\\[2mm]
&&
c^{-}_{-}(c,l,u_0)=-c^{+}_{+}(c,l,u_0),\quad c^{-}_{+}(c,l,u_0)=-c^{+}_{-}(c,l,u_0).
\ena

Comparing ${\bar K}_g(u_g;\xi_g,\lambda_g,\mu_g)$ with ${\bar K}(u;c,l,u_0)$, we define
\bea
&&
\mu'=\mu'(l,c,u_0)\non\\
&&
\quad =\frac{c^{+}_{-}(c,l,u_0)c^{\enskip\, +}_{g,+}(\xi'(c,l,u_0))}{c^{+}_{+}(c,l,u_0)c^{\enskip\, +}_{g,-}(\xi'(c,l,u_0),a'(c,l,u_0))}.
\ena
then we have the desired result
\bea
&&
\frac{c^{+}_{+}(c,l,u_0)}{c^{\enskip\, +}_{g,+}(\xi'(c,l,u_0))}
{\bar K}_g(u;\xi',a',\mu')={\bar K}(u;c,l,u_0)
\ena
where we abuse the notation for ${\bar K}_g$
\bea
{\bar K}_g(u;\xi,a,\mu_g)={\bar K}_g(u_g;\xi_g,\lambda_g,\mu_g).
\ena

%
%
 
\pagebreak

\newpage

\begin{figure}
 \begin{center}
  \includegraphics{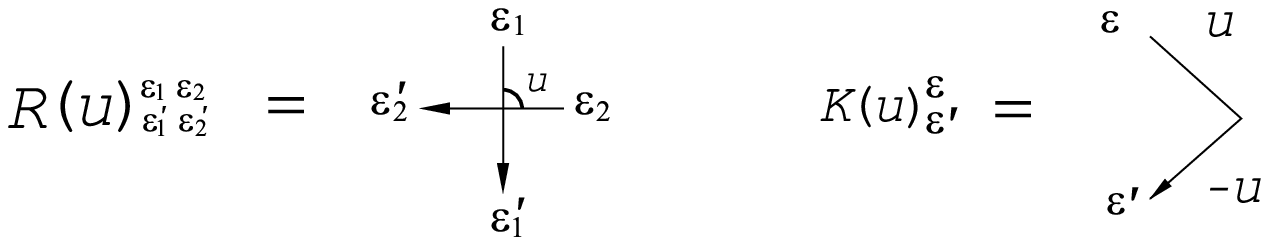}
  \caption{Graphical representation of $R$ and $K$-matrix.}
 \lb{fig:R-K}
 \end{center}
\end{figure}

\begin{figure}
 \begin{center}
  \includegraphics{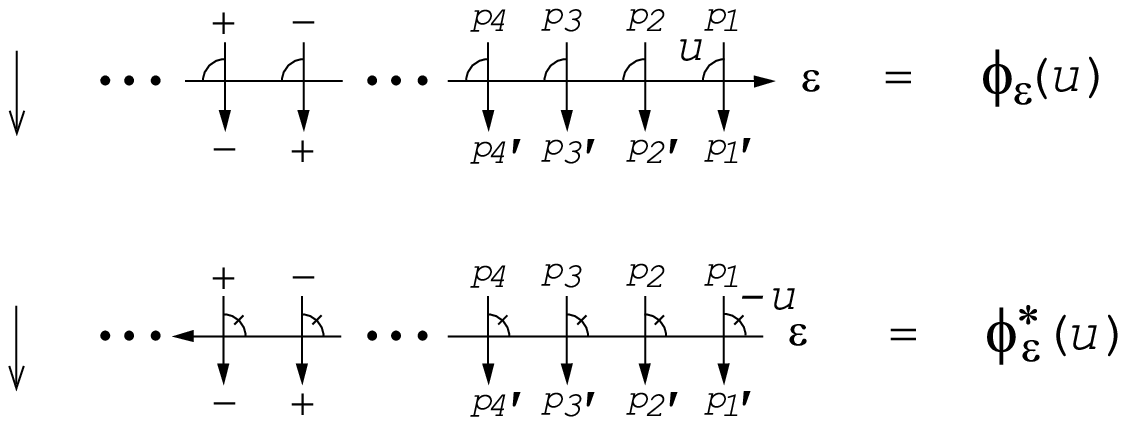}
  \caption{Lattice vertex operators of vertex type}
  \lb{fig:VO-v}
 \end{center}
\end{figure}

\begin{figure}
 \begin{center}
  \includegraphics{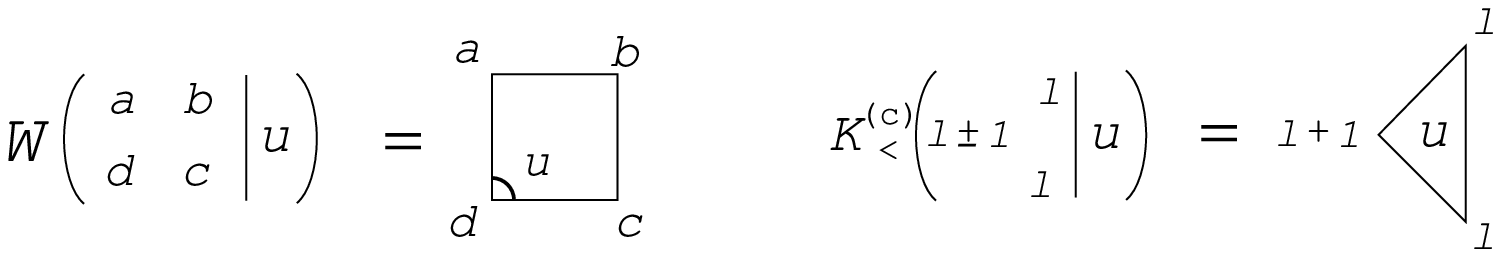}
  \caption{Graphical representation of the bulk and boundary Boltzmann weights of the SOS model.}
 \lb{fig:W-K}
 \end{center}
\end{figure}

\begin{figure}
 \begin{center}
  \includegraphics{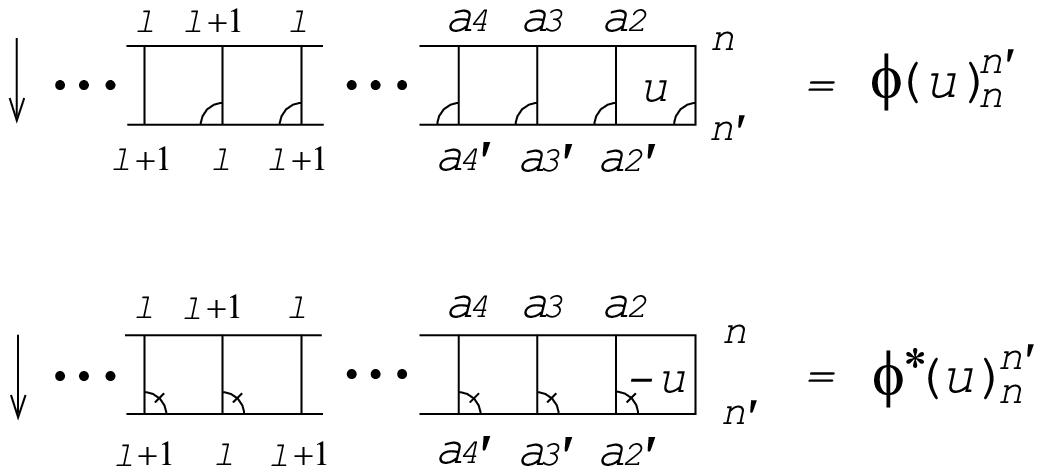}
  \caption{Lattice vertex operators of face type.}
  \lb{fig:VO-f}
 \end{center}
\end{figure}


\begin{figure}
 \begin{center}
  \includegraphics{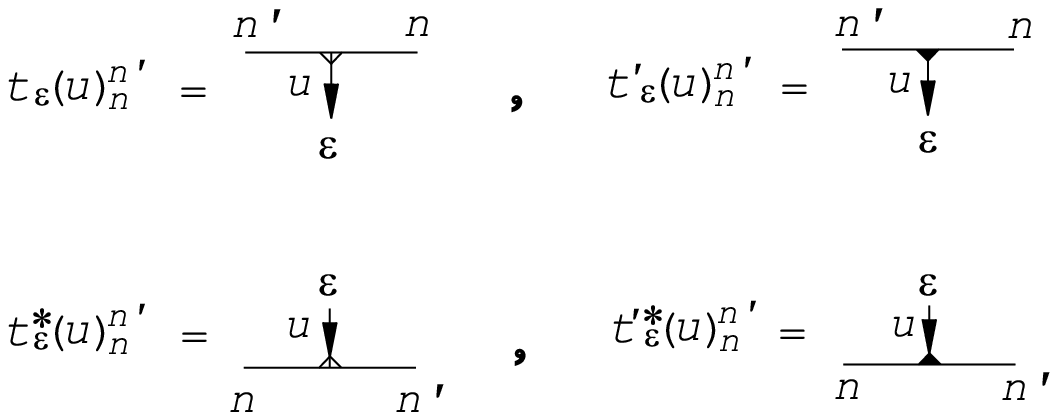}
  \caption{Graphical representation of intertwining vectors.}
  \lb{fig:vector}
 \end{center}
\end{figure}
 
\begin{figure}
 \begin{center}
  \includegraphics{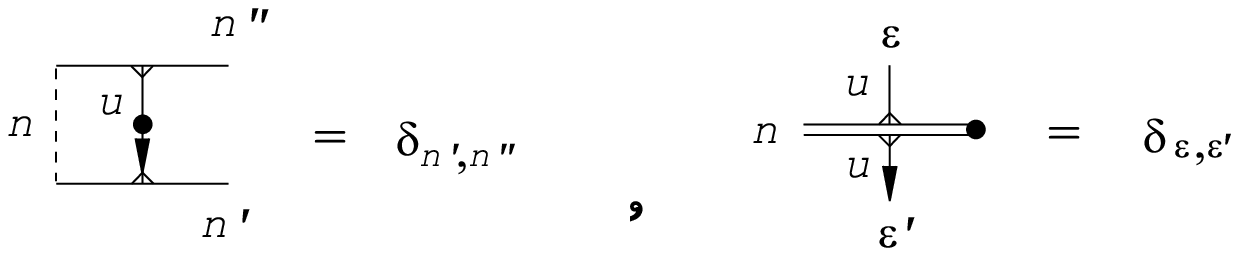}
  \caption{Graphical representation of \eqref{eqn:vec-rel}.}
  \lb{fig:vec-rel}
 \end{center}
\end{figure}

\begin{figure}
 \begin{center}
  \includegraphics{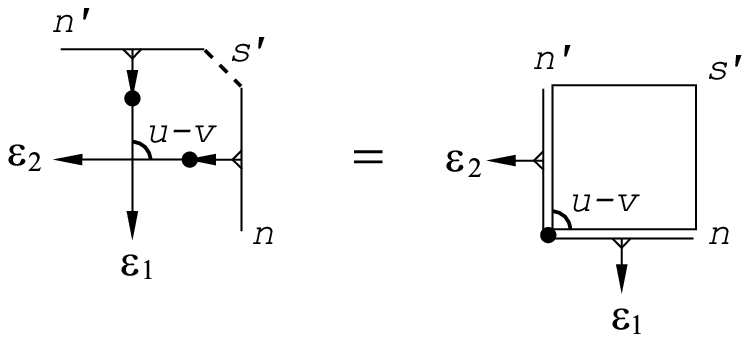}
  \caption{The face-vertex correspondence of bulk weights \eqref{eqn:fv-RW}.}
  \lb{fig:fv-RW}
 \end{center}
\end{figure}

\begin{figure}
 \begin{center}
  \includegraphics{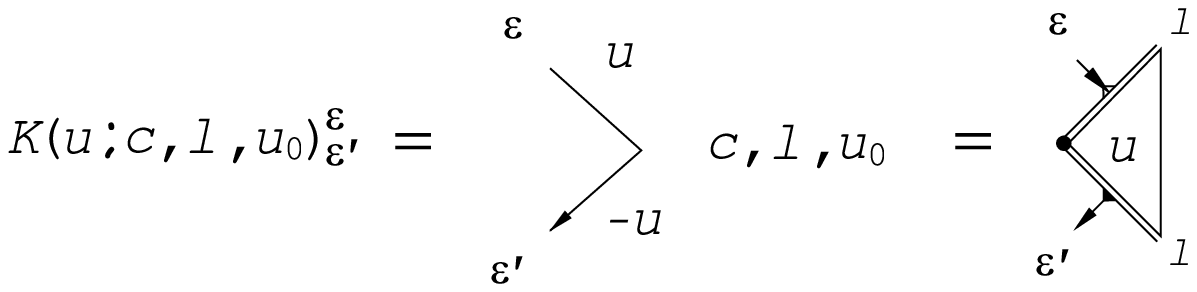}
  \caption{The face-vertex correspondence of $K$-matrix \eqref{eqn:fv-K}.}
  \lb{fig:fv-K}
 \end{center}
\end{figure}

\begin{figure}
 \begin{center}
  \includegraphics{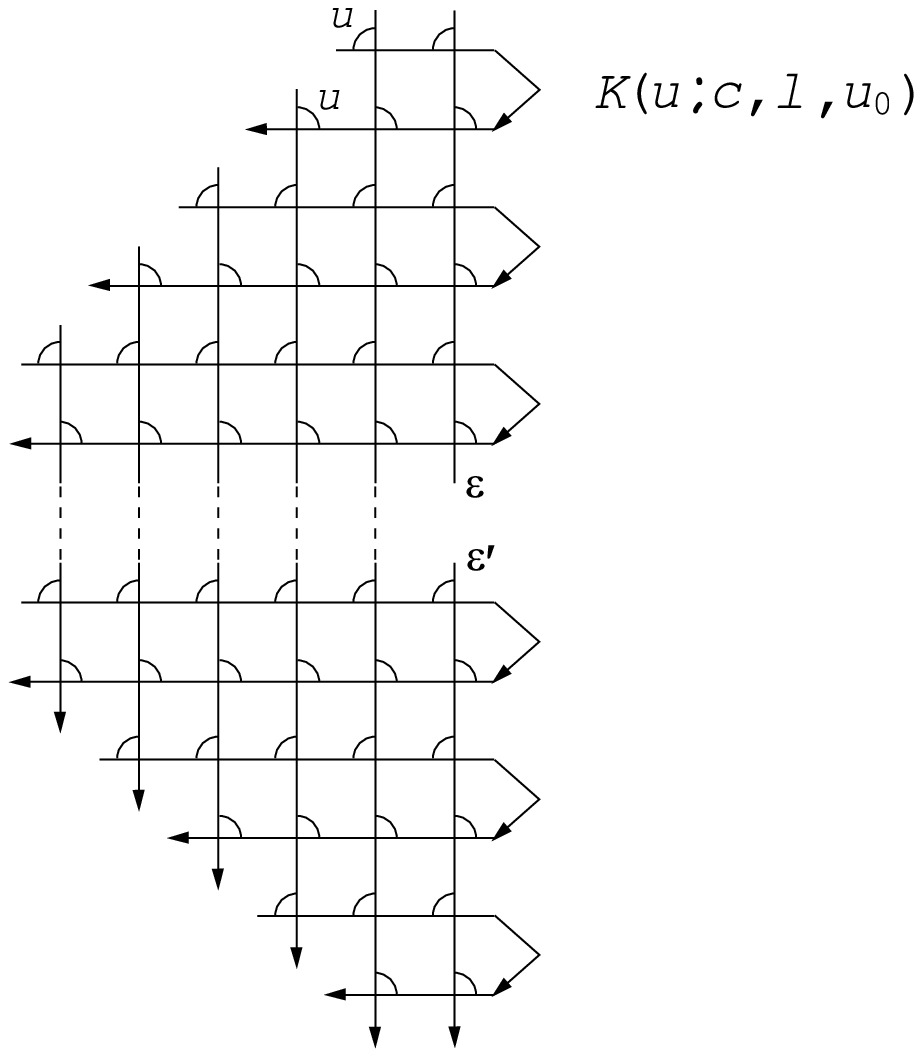}
  \caption{The boundary eight-vertex model. Lattice extends to the north, the west and the south.}
  \lb{fig:latt-v}
 \end{center}
\end{figure}

\begin{figure}
 \begin{center}
  \includegraphics{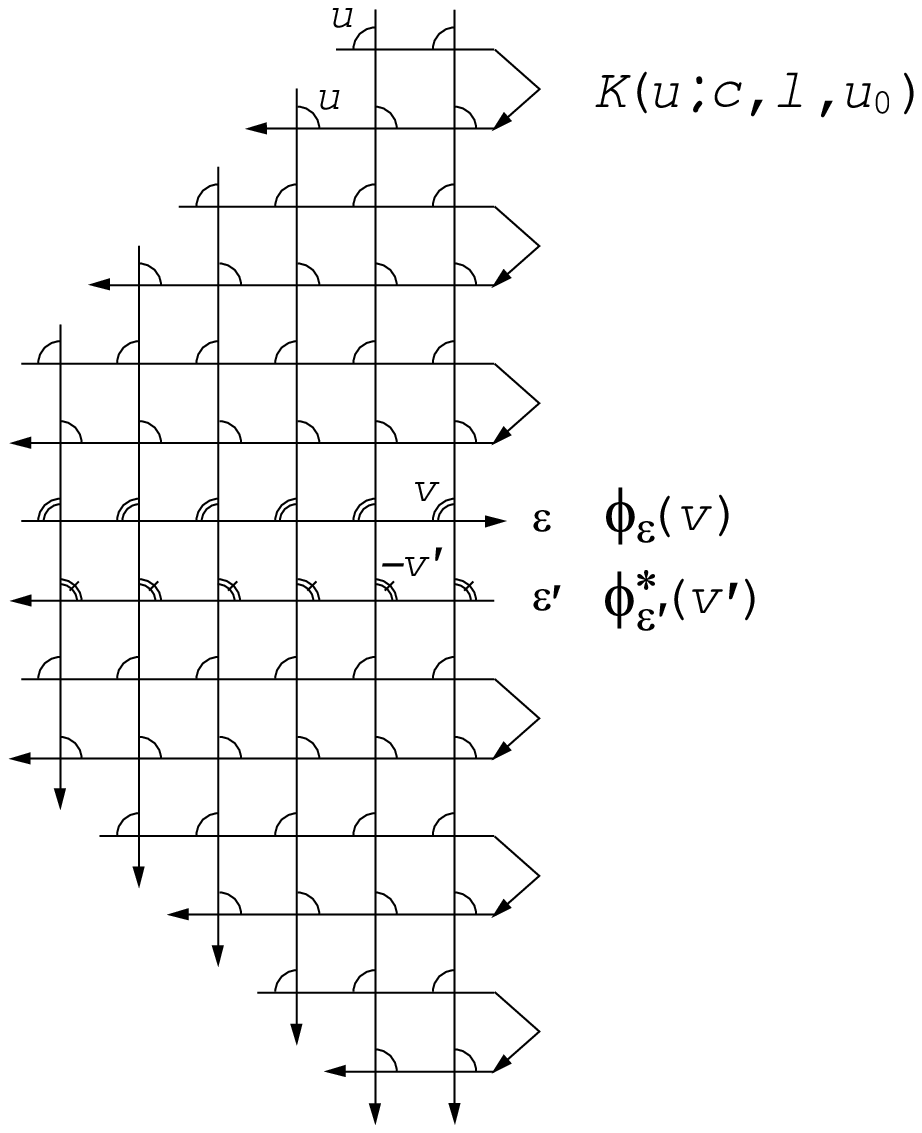}
  \caption{The boundary eight-vertex model corresponding to the boundary magnetization.}
  \lb{fig:ctm-v}
 \end{center}
\end{figure}

\begin{figure}
 \begin{center}
  \includegraphics{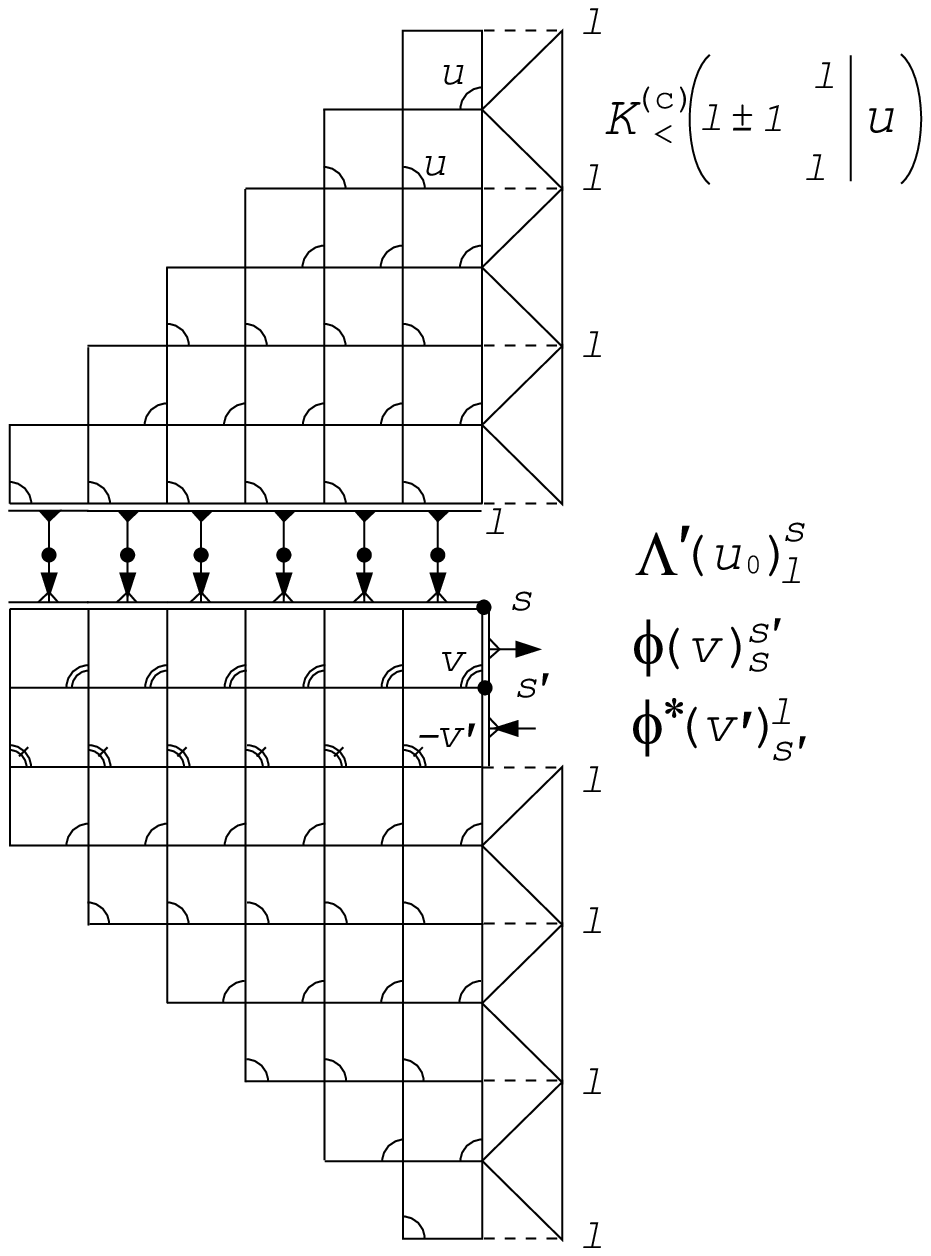}
  \caption{The boundary SOS model corresponding to the boundary magnetization. Summations including intertwining vectors are emphasized with black dots for clarity.}
  \lb{fig:ctm-f}
 \end{center}
\end{figure}

\begin{figure}
 \begin{center}
  \includegraphics{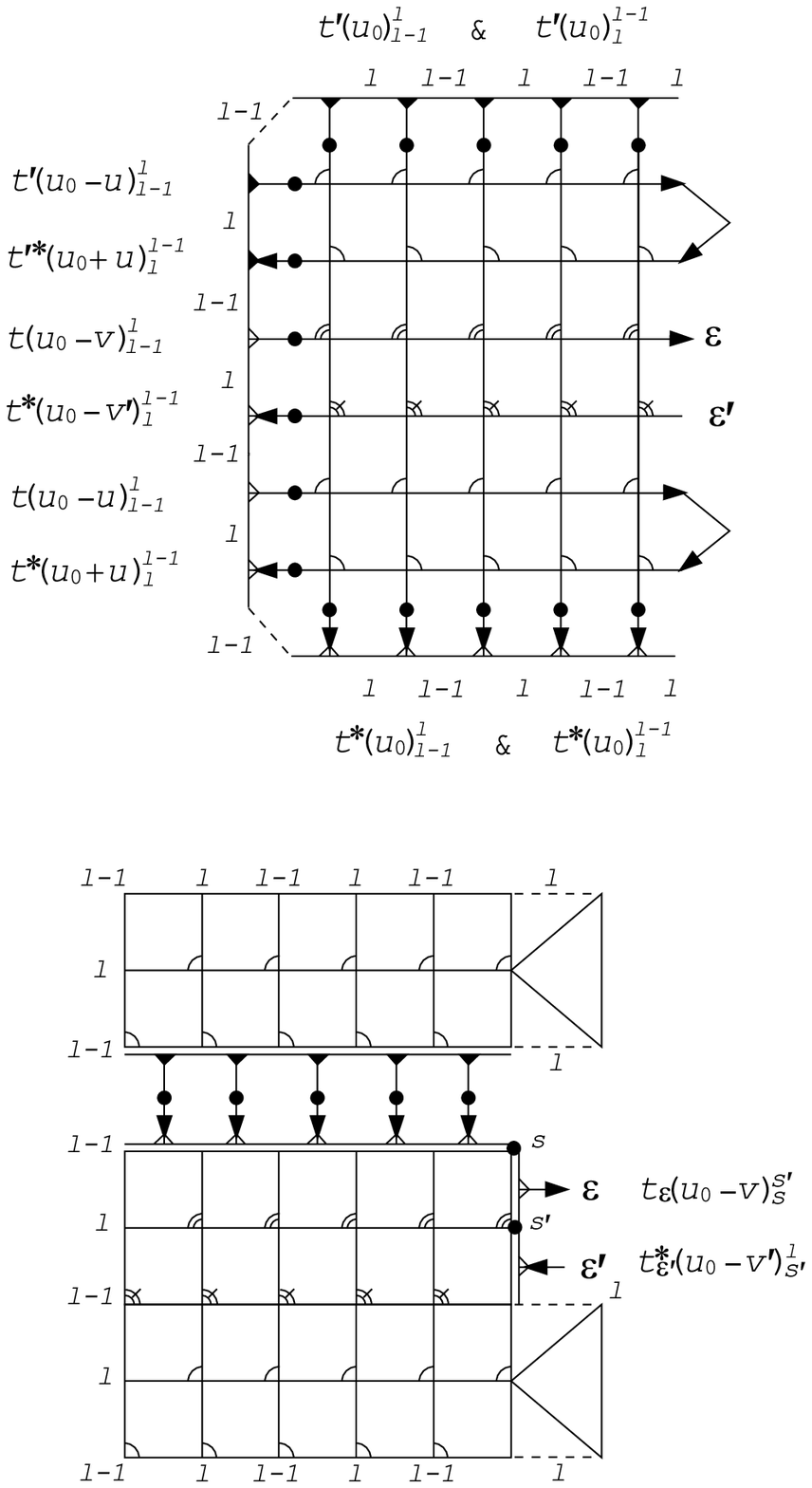}
  \caption{Face-vertex correspondence of finite size lattice.}
  \lb{fig:fv-latt}
 \end{center}
\end{figure}

\end{document}